\documentclass[journal]{IEEEtran}

\usepackage{cite} 

\ifCLASSINFOpdf
  \usepackage[pdftex]{graphicx}
  \graphicspath{{images/}}  
  \DeclareGraphicsExtensions{.pdf,.jpeg,.png}
\else 
  \usepackage[dvips]{graphicx}
  \graphicspath{{images/}}
  \DeclareGraphicsExtensions{.eps}
\fi  
  
\usepackage[cmex10]{amsmath}
\usepackage{amssymb}
\usepackage{amsthm}

\usepackage{amsbsy}
\usepackage[mathcal]{eucal}

\usepackage{algorithm}
\usepackage{algpseudocode} 
\usepackage{xspace}
\algnewcommand{\algorithmicgoto}{\textbf{go to}}%
\algnewcommand{\Goto}{\algorithmicgoto\xspace}%
\algnewcommand{\Label}{\State\unskip}
\algnewcommand{\Break}{\State \textbf{break}}

\usepackage{tabularx}

\usepackage{array}

\usepackage{mathrsfs}
\DeclareMathAlphabet{\mathpzc}{OT1}{pzc}{m}{it}
\usepackage[caption=false,font=footnotesize]{subfig}

\usepackage{url}

\usepackage{fancybox}

\usepackage{calc}
\newsavebox\myboxa

\usepackage{overpic}
\usepackage{psfrag}

\usepackage{dblfloatfix}

\usepackage{txfonts}
\usepackage{textcomp}

\usepackage{stmaryrd}

\usepackage{pbox}

\usepackage{diagbox}

\usepackage{xcolor}
\usepackage{colortbl}

\usepackage{hhline}

\usepackage{multirow}

\usepackage{makecell}

\usepackage{pifont}
\usepackage{soul}

\usepackage[export]{adjustbox}

\usepackage{enumitem}

\usepackage{bbding}

\makeatletter
\newsavebox\saved@arstrutbox
\newcommand*{\setarstrut}[1]{%
  \noalign{%
    \begingroup
      \global\setbox\saved@arstrutbox\copy\@arstrutbox
      #1%
      \global\setbox\@arstrutbox\hbox{%
        \vrule \@height\arraystretch\ht\strutbox
               \@depth\arraystretch \dp\strutbox
               \@width\z@
      }%
    \endgroup
  }%
}
\newcommand*{\restorearstrut}{%
  \noalign{%
    \global\setbox\@arstrutbox\copy\saved@arstrutbox
  }%
}
\makeatother

\makeatletter
\newcommand{\algmultiline}[1]{%
  \begin{tabularx}{\dimexpr\linewidth-\ALG@thistlm}[t]{@{}X@{}}
    #1
  \end{tabularx}
}
\makeatother

\newcommand{\argmin}{\operatornamewithlimits{argmin}}

\makeatletter

\newcommand{\Rmnum}[1]{\expandafter\@slowromancap\romannumeral #1@}
\makeatother

\begin{document}

\title{CALPA-NET: Channel-pruning-assisted Deep Residual Network for
  Steganalysis of Digital Images}

\author{ Shunquan~Tan,~\IEEEmembership{Senior Member,~IEEE,}
  Weilong~Wu, Zilong~Shao, Qiushi~Li,
  Bin~Li,~\IEEEmembership{Senior Member,~IEEE,}
  and~Jiwu~Huang,~\IEEEmembership{Fellow,~IEEE}%

  \thanks{S. Tan, W. Wu, Z. Shao, and Q. Li are with College of
    Computer Science and Software Engineering, Shenzhen
    University. B. Li and J. Huang are with College of Electronic and
    Information Engineering, Shenzhen University.}  

  \thanks{All of the members are with the Guangdong Key Laboratory of
    Intelligent Information Processing, Shenzhen Key Laboratory of
    Media Security, Guangdong Laboratory of Artificial Intelligence
    and Digital Economy (SZ), Shenzhen Institute of Artificial
    Intelligence and Robotics for Society, China (email:
    tansq@szu.edu.cn).}

  \thanks{This work was supported in part by the Key-Area Research and
    Development Program of Guangdong Province~(2019B010139003),
    NSFC~(61772349, U19B2022, U1636202, 61872244), Guangdong Basic and
    Applied Basic Research Foundation~(2019B151502001), and Shenzhen
    R\&D Program~(GJHZ20180928155814437, JCYJ20180305124325555). This
    work was also supported by Alibaba Group through Alibaba
    Innovative Research (AIR) Program.}

  \thanks{This paper has supplementary downloadable material available
    at http://ieeexplore.ieee.org, provided by the author. The
    material includes source codes and analysis results. Contact
    tansq@szu.edu.cn for further questions about this work.}%
}
\maketitle
\begin{abstract}
  Over the past few years, detection performance improvements of
  deep-learning based steganalyzers have been usually achieved through
  structure expansion. However, excessive expanded structure results
  in huge computational cost, storage overheads, and consequently
  difficulty in training and deployment. In this paper we propose
  CALPA-NET, a ChAnneL-Pruning-Assisted deep residual network
  architecture search approach to shrink the network structure of
  existing vast, over-parameterized deep-learning based
  steganalyzers. We observe that the broad inverted-pyramid structure
  of existing deep-learning based steganalyzers might contradict the
  well-established model diversity oriented philosophy, and therefore
  is not suitable for steganalysis. Then a hybrid criterion combined
  with two network pruning schemes is introduced to adaptively shrink
  every involved convolutional layer in a data-driven manner. The
  resulting network architecture presents a slender bottleneck-like
  structure. We have conducted extensive experiments on BOSSBase+BOWS2
  dataset, more diverse ALASKA dataset and even a large-scale subset
  extracted from ImageNet CLS-LOC dataset. The experimental results
  show that the model structure generated by our proposed CALPA-NET
  can achieve comparative performance with less than two percent of
  parameters and about one third FLOPs compared to the original
  steganalytic model. The new model possesses even better adaptivity,
  transferability, and scalability.
\end{abstract}


% Define bibtex format. e.g. the usage of "et al."
\bstctlcite{psm_bstctl}

\section{Introduction}
\label{sec:intro}

\IEEEPARstart{S}{teganalysis} aims to detect covert communication
established via steganography. In addition to detection performance,
computational cost, model complexity, as well as adaptivity,
transferability, and scalability, are all important considerations for
a real-world steganalytic framework.

% steganography
Over the last decade, the main battleground of the war between modern
steganography and steganalysis has always been in digital
images~\cite{bohme_ast_ver1_2010}. Most of the spatial-domain and
frequency-domain image steganographic algorithms have adopted the
embedding distortion minimizing framework~\cite{filler_tifs_2010}. The
prominent additive embedding distortion minimizing schemes include
HILL~\cite{li_icip_2014} and MiPOD~\cite{sedighi_tifs_2016} in spatial
domain, as well as UERD~\cite{guo_tifs_2015} in JPEG
domain. UNIWARD~\cite{holub_eurasip_2014} is a benchmarking
cross-domain scheme~(noted as S-UNIWARD in spatial domain, while
J-UNIWARD in JPEG domain).

In recent years, deep-learning frameworks, especially Convolutional
Neural Networks~(CNNs) have achieved overwhelming
superiority~\cite{schmidhuber_nn_2015}. In the meanwhile, from early
AlexNet~\cite{krizhevsky_nips_2012} and
VGGNet~\cite{simonyan_arxiv_1409.1556}, to later Inception
models~\cite{szegedy_cvpr_2015} and ResNet~\cite{he_cvpr_2016}, the
literature witnessed deep-learning frameworks have become more and
more deeper and complicated.

% traditional steganalytic features
The once overlord of image steganalysis is the ``rich model''
hand-crafted features
family~\cite{fridrich_tifs_2012,denemark_wifs_2014,tang_tifs_2016,tan_tifs_2017,li_tifs_2018}
equipped with an ensemble
classifier~\cite{kodovsky_tifs_2012}. Started from the work of Tan and
Li~\cite{tan_apsipa_2014}, breakthroughs were made in deep-learning
based
steganalysis~\cite{qian_spie_2015,pibre_ei_2016_ver2,xu_spl_2016}. Then
Ye et al. proposed a deep-learning based steganalyzer equipped with a
new activation function called Truncated Linear
Unit~(TLU)~\cite{ye_tifs_2017}, achieving significant improvement in
spatial domain. In JPEG domain, Chen et al. proposed a specific
deep-learning based steganalyzer aware of JPEG
phase~\cite{chen_ihmmsec_2017}. Xu proposed a novel 20-layer framework
with residual connections~(named XuNet2)~\cite{xu_ihmmsec_2017}.
Aiming at large-scale JPEG image steganalysis, Zeng et al. proposed a
generic hybrid deep-learning framework incorporating the domain
knowledge behind rich steganalytic models~\cite{zeng_tifs_2018}. On
the contrary, Boroumand et al. proposed a deep residual steganalytic
network designed to minimize the use of heuristic domain
knowledge~(named SRNet)~\cite{boroumand_tifs_2018}, which has shown
superior detection performance for both spatial domain and JPEG domain
steganography. In \cite{zeng_tifs_2019}, Zeng et al. explicitly
considered the correlation among color bands and proposed WISERNet,
the wider separate-then-reunion network specifically designed for
steganalysis of true-color images. For a comprehensive survey please
refer to \cite{chaumont_arxiv_1904.01444}.

It has been widely acknowledged that existing deep-learning frameworks
are over parameterized, and are therefore with huge computational cost
and storage overheads. Consequently, they are more likely to require
massive amounts of training data to achieve good performance, and are
hard to deploy~\cite{howard_arxiv_1704.04861}.  As also discussed in
\cite{chaumont_arxiv_1904.01444}, automatic network architecture
searching techniques~\cite{liu_eccv_2018} might be potential to build
up an effective deep-learning framework.  Another trend is network
pruning, which is one of the most popular methods to reduce network
complexity. Network pruning has become one important research problem
even since a very early stage of the evolution of deep-learning
frameworks~\cite{lecun_nips_1990}. For CNNs, state-of-the-art network
pruning approaches can be classified into two categories, the
non-structured weights pruning and the channel-based structured
pruning methods. Non-structured weights pruning
methods~\cite{han_nips_2015} cannot lead to complexity reduction
without the support of dedicated hardware. Therefore structured
pruning methods are more practical. Hu et al. proposed a channel
pruning method based on the percentage of zeros in the
outputs~\cite{hu_arxiv_1607.03250}. Li et al. proposed another channel
pruning method based on the $l_1$ norms of the corresponding filter
weights~\cite{li_iclr_2017}. Luo et al. proposed ThiNet which greedily
prunes those channels with smallest effects on the activation values
of the next layer~\cite{luo_iccv_2017}. Further on, Huang and Wang
used sparsity regularization to prune channels and even coarser
structures, such as residual blocks~\cite{huang_eccv_2018}. The works
mentioned above are just the notable ones selected from a complete
collection. But, whatever the approach, the existing network pruning
approaches have all followed a typical three-stage pipeline: training,
pruning, and then finetuning in order to preserve a set of inherited
learned important weights.

In this paper, we propose CALPA-NET, a channel-pruning-assisted deep
residual network architecture search approach to shrink the network
structure of existing vast, over-parameterized deep-learning based
steganalyzers. We observe that the established ``doubling the number
of channels along with halving the size of output feature maps'' rule
might undermine the diversity of output features of deep steganalytic
models. Therefore starting from existing bloated deep steganalytic
models, CALPA-NET utilizes a hybrid criterion to adaptively determine
the number of channels of every involved convolutional layer. The
proposed hybrid criterion combines the ThiNet
scheme~\cite{luo_iccv_2017} and the $l_1$-norm based
scheme~\cite{li_iclr_2017}. Please note that CALPA-NET, where we
abandon the training-pruning-finetuning pipeline of typical network
pruning methods, is entirely different from network pruning approaches
mentioned above since the proposed channel pruning criterion is just
used to search efficient network architectures.  The extensive
experiments conducted on public datasets show that even trained from
scratch, the shrunken model can still achieve state-of-the-art
detection performance with merely a tiny proportion of parameters and
much lower computational complexity.

The rest of the paper is organized as
follows. Sect.~\ref{sec:proposed} firstly gives a brief overview of
existing representative deep steganalytic models as well as
channel-based structured pruning methods. Then experimental
testimonies are provided to support the rationale of CALPA-NET. Next
the procedure with our proposed hybrid channel pruning criterion used
in CALPA-NET is described in detail. Results of experiments conducted
on public datasets are presented in Sect.~\ref{sec:exp}. Finally, we
make a conclusion in Sect.~\ref{sec:conclude}.

\begin{figure*}[!t]
  \centering
  \includegraphics[width=1.1\textwidth,keepaspectratio,center]{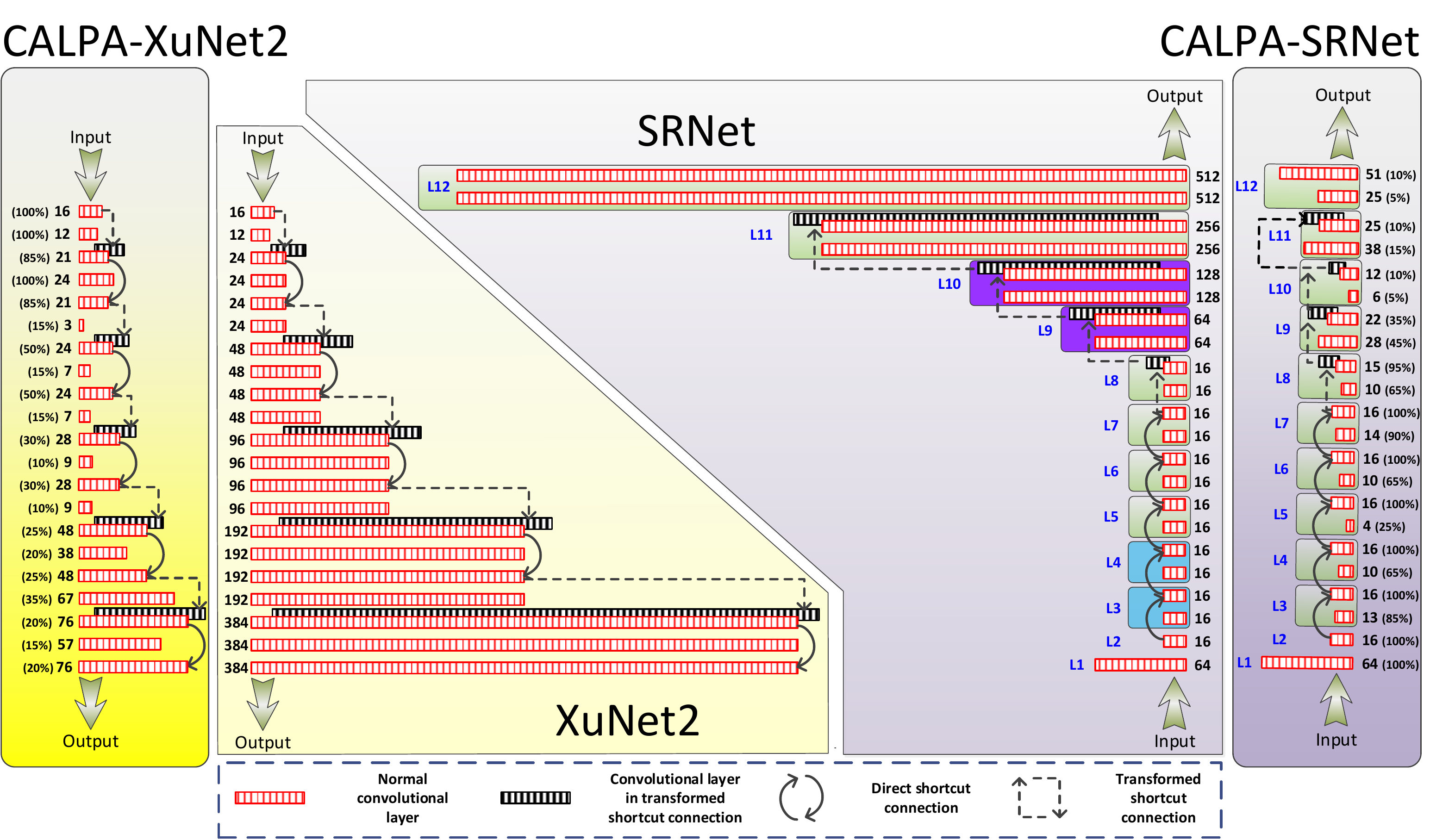}
  \caption[]{From left to right, the conceptual structures of 
    CALPA-XuNet2, XuNet2, SRNet, and CALPA-SRNet\textsuperscript{*}
    are illustrated respectively. The number and corresponding
    shrinking rate of output channels of every convolutional layer is
    shown alongside the representing bar. For SRNet and CALPA-SRNet,
    blue ``L1'' to ``L12'' represent twelve composition blocks
    following the notations in \cite{boroumand_tifs_2018}.}
  {\scriptsize\textsuperscript{*} The shrinking rates of CALPA-XuNet2 and
  CALPA-SRNet are aiming at detecting J-UNIWARD stego images with 0.4
  bpnzAC payload and QF 75. }
  \label{fig:calpa_xunet_srnet_structure}  
\end{figure*} 

\section{Our proposed CALPA-NET}
\label{sec:proposed}

\subsection{Preliminaries}
\label{sec:pre}

As far as we know, existing deep-learning based steganalyzers are all
based on CNNs~(Convolutional Neural Network). The principal part of
CNN can be modeled as a direct graph of alternating convolutional
layers and auxiliary layers~(e.g. BN~(Batch Normalization)
layers and pooling layers). Convolutional
layers are the central components of a CNN since they contain the
overwhelming majority of learnable weights and biases. For a given
convolutional layer $L_l$, it takes an after-activation input tensor
$\boldsymbol{\widehat{\mathcal{Z}}}^{l-1} \in \mathbb{R}^{J^{l-1}
  \times H^{l-1} \times W^{l-1}}$ which possesses $J^{l-1}$ input
channels with height $H^{l-1}$ and width $W^{l-1}$, convolves it with
$\boldsymbol{\mathcal{W}}^{l} \in \mathbb{R}^{J^{l-1} \times K^{l}
  \times W^{l} \times W^{l}}$, a filter tensor consisting of
$J^{l-1} \times K^{l}$ kernels with size $W^{l} \times W^{l}$, and
generates
$\boldsymbol{\mathcal{Z}}^{l} \in \mathbb{R}^{K^{l} \times H^{l}
  \times W^{l}}$, the corresponding before-activation output tensor
which has $K^{l}$ output channels~(feature maps) with height $H^{l}$
and width $W^{l}$. The convolution operation can be modeled as~(the
bias is omitted for brevity):
\begin{equation}
  \label{eq:conv}
  \boldsymbol{\mathcal{Z}}^{l}_{k::}=
  \sum^{J^{l-1}}_{j=1}\boldsymbol{\widehat{\mathcal{Z}}}^{l-1}_{j::} \ast \boldsymbol{\mathcal{W}}^{l}_{jk::},\ 1 \le k \le K^{l}
\end{equation}
where $\boldsymbol{\widehat{\mathcal{Z}}}^{l-1}_{j::}$,
$\boldsymbol{\mathcal{W}}^{l}_{jk::}$, and
$\boldsymbol{\mathcal{Z}}^{l}_{k::}$ denotes the \textit{slides}, the
two dimensional sections defined by fixing all but the last two
indices, of $\boldsymbol{\widehat{\mathcal{Z}}}^{l-1}$,
$\boldsymbol{\mathcal{W}}^{l}$, and $\boldsymbol{\mathcal{Z}}^{l}$,
respectively.

\subsubsection{Interior structures of deep-learning based steganalyzers}
\label{sec:pre-deep-learning-steganalyzers}

Two popular deep-learning based steganalyzers,
SRNet~\cite{boroumand_tifs_2018} and XuNet2~\cite{xu_ihmmsec_2017} are
taken as examples. Their conceptual structures are illustrated in
Fig.~\ref{fig:calpa_xunet_srnet_structure}. The two deep-learning
based steganalyzers both take spatial representation of the target
image as input.  All deep-learning based steganalyzers can be divided
into three modules: the bottom module aiming at extracting the
so-called ``stego noise residuals'', the middle module which striving
for learning compact representative features and the top module which
is a simple binary classifier.

Actually most of the existing research works regarding to
deep-learning based steganalyzers have been devoted to the bottom
module. The bottom module of XuNet2 adopts a particular truncated
filter bank with sixteen $4 \times 4$ DCT high-pass filters. Further
on, the bottom module of SRNet consists of a pile of two hierarchical
convolutional layers~(``L1'' and ``L2'', following the notations in
\cite{boroumand_tifs_2018}) and five unpooled residuals blocks with
direct shortcut connections~(from ``L3'' to ``L7'') for extraction of
noise residuals. Conversely, according to our best knowledge, no
specific domain knowledge has ever been introduced to guide the design
of the middle module as well as the top module. Researchers just
simply followed the effective recipes in other research
fields~(e.g. computer vision). Nowadays, almost all of the
state-of-the-art deep-learning based steganalyzers, including SRNet
and XuNet2, have adopted shortcut connections inspired by
ResNet~\cite{he_cvpr_2016} and a simple design rule--- ``doubling the
number of channels along with halving the spatial size of feature
maps'' which can be traced back to
VGGNet~\cite{simonyan_arxiv_1409.1556}. Take SRNet for instance. As
shown in Fig.~\ref{fig:calpa_xunet_srnet_structure}, from ``L3'' up to
``L11'' are with shortcut connections. From its ``L9'' up to ``L12'',
the doubling numbers of output channels~(64, 128, 256, and 512)
correspond respectively to the halving sizes of feature
maps~($128 \times 128$, $64 \times 64$, $32 \times 32$, and
$16 \times 16$).

\subsubsection{Channel pruning methods: ThiNet and $l_1$-norm based}
\label{sec:pre-channel-pruning-methods}

Since the training-pruning-finetuning pipeline of typical network
pruning methods is completed abandoned in CALPA-NET, only the pruning
criteria in ThiNet~\cite{luo_iccv_2017} and the $l_1$-norm based
scheme~\cite{li_iclr_2017} are addressed.

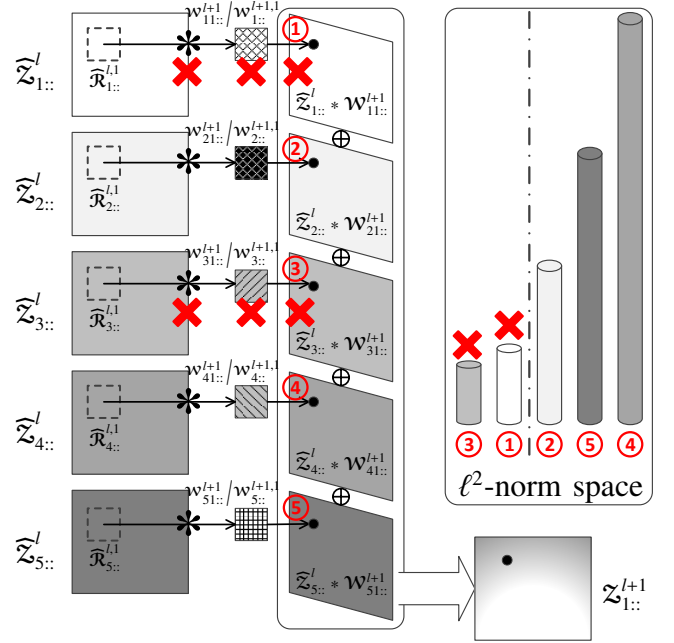
\begin{figure}[!t]
  \centering
%  \includegraphics[width=\linewidth,keepaspectratio,center]{eq1_eq2_illustration}
%  \begin{overpic}[width=\linewidth,keepaspectratio,grid,tics=4]{eq1_eq2_illustration}
  \begin{overpic}[width=\linewidth,keepaspectratio]{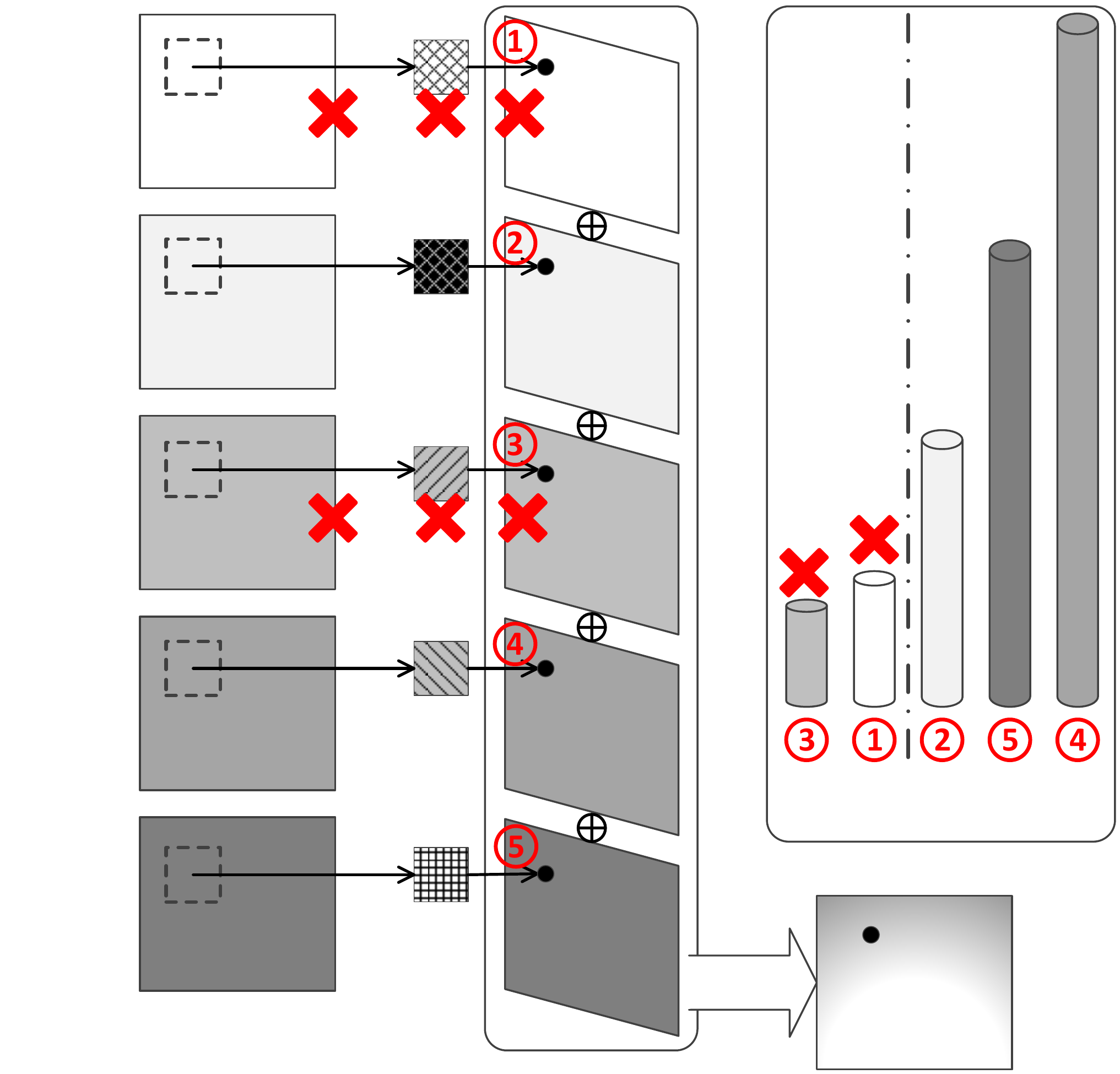}
    \put(92,6){$\boldsymbol{\mathcal{Z}}^{l+1}_{1::}$}
    \put(4,84){$\boldsymbol{\widehat{\mathcal{Z}}}^{l}_{1::}$}
    \put(4,66){$\boldsymbol{\widehat{\mathcal{Z}}}^{l}_{2::}$}
    \put(4,48){$\boldsymbol{\widehat{\mathcal{Z}}}^{l}_{3::}$}
    \put(4,30){$\boldsymbol{\widehat{\mathcal{Z}}}^{l}_{4::}$}
    \put(4,12){$\boldsymbol{\widehat{\mathcal{Z}}}^{l}_{5::}$}
    \put(29.2,94){\tiny$\boldsymbol{\mathcal{W}}^{l+1}_{11::}\left/\boldsymbol{\mathcal{W}}^{l+1,1}_{1::}\right.$}
    \put(29.2,76.1){\tiny$\boldsymbol{\mathcal{W}}^{l+1}_{21::}\left/\boldsymbol{\mathcal{W}}^{l+1,1}_{2::}\right.$}
    \put(29.2,57.8){\tiny$\boldsymbol{\mathcal{W}}^{l+1}_{31::}\left/\boldsymbol{\mathcal{W}}^{l+1,1}_{3::}\right.$}
    \put(29.2,40.2){\tiny$\boldsymbol{\mathcal{W}}^{l+1}_{41::}\left/\boldsymbol{\mathcal{W}}^{l+1,1}_{4::}\right.$}
    \put(29.2,22){\tiny$\boldsymbol{\mathcal{W}}^{l+1}_{51::}\left/\boldsymbol{\mathcal{W}}^{l+1,1}_{5::}\right.$}
   \put(15,83.5){\scriptsize$\boldsymbol{\widehat{\mathcal{R}}}^{l,1}_{1::}$} 
   \put(15,65.5){\scriptsize$\boldsymbol{\widehat{\mathcal{R}}}^{l,1}_{2::}$} 
   \put(15,47.5){\scriptsize$\boldsymbol{\widehat{\mathcal{R}}}^{l,1}_{3::}$} 
   \put(15,29.5){\scriptsize$\boldsymbol{\widehat{\mathcal{R}}}^{l,1}_{4::}$} 
   \put(15,11.5){\scriptsize$\boldsymbol{\widehat{\mathcal{R}}}^{l,1}_{5::}$} 
    \put(28,88.2){\Asterisk}
    \put(28,70.4){\Asterisk}
    \put(28,52.1){\Asterisk}
    \put(28,34.3){\Asterisk}
    \put(28,16){\Asterisk}
    \put(46,80){\scriptsize$\boldsymbol{\widehat{\mathcal{Z}}}^{l}_{1::}
      \ast \boldsymbol{\mathcal{W}}^{l+1}_{11::}$}
    \put(46,62){\scriptsize$\boldsymbol{\widehat{\mathcal{Z}}}^{l}_{2::}
      \ast \boldsymbol{\mathcal{W}}^{l+1}_{21::}$}
    \put(46,44){\scriptsize$\boldsymbol{\widehat{\mathcal{Z}}}^{l}_{3::}
      \ast \boldsymbol{\mathcal{W}}^{l+1}_{31::}$}
    \put(46,26){\scriptsize$\boldsymbol{\widehat{\mathcal{Z}}}^{l}_{4::}
      \ast \boldsymbol{\mathcal{W}}^{l+1}_{41::}$}
    \put(46,8){\scriptsize$\boldsymbol{\widehat{\mathcal{Z}}}^{l}_{5::} 
      \ast \boldsymbol{\mathcal{W}}^{l+1}_{51::}$}
   %  \put(76,87){$\color{red}\scriptsize \left( \begin{array}{c}
   %                                               \boldsymbol{\widehat{\mathcal{Z}}}^{l}_{1::}\ast\boldsymbol{\mathcal{W}}^{l+1,1}_{1::}
   %                                               \\
   %                                               +
   %                                               \\
   %                                               \boldsymbol{\widehat{\mathcal{Z}}}^{l}_{3::}\ast\boldsymbol{\mathcal{W}}^{l+1,1}_{3::}
   %                                             \end{array}
   %                                           \right)^2$}
   % \put(84,74){\textcolor{red}{ \LARGE \bf \textless} }                                                                                      
   %  \put(76,62){$\color{red}\scriptsize \left( \begin{array}{c}
   %                                               \boldsymbol{\widehat{\mathcal{Z}}}^{l}_{i::}\ast\boldsymbol{\mathcal{W}}^{l+1,1}_{i::}
   %                                               \\
   %                                               +
   %                                               \\
   %                                               \boldsymbol{\widehat{\mathcal{Z}}}^{l}_{j::}\ast\boldsymbol{\mathcal{W}}^{l+1,1}_{j::}
   %                                             \end{array}
   %                                           \right)^2$}
   %  \put(76,50){$\color{red}\scriptsize\begin{array}{c}i<j\\ \ i,j \in \{2,4,5\}\end{array}$}                                           
    \put(70.5,22){\large $\ell^2$-norm space}
  \end{overpic}  
  \caption[]{Illustration of ThiNet
    algorithm~(Eq.~\eqref{eq:thinet_optimization_problem}) via a toy example.}
  \label{fig:eq1_eq2_illustration}  
\end{figure} 

ThiNet~\cite{luo_iccv_2017} is a data-driven channel selection
algorithm. For a pre-defined pruning rate $\gamma$, a subset of $m$
images are randomly selected from training set and fed to a trained
CNN model one by one. Fed with every image in the subset, the input
and output tensors of all the convolutional layers of the CNN model
are obtained in a forward propagation.  In ThiNet, in order to prune a
given convolutional layer $L_{l}$, we have to observe its impact on
the next layer $L_{l+1}$.  Given a specific convolutional layer
$L_{l+1}$, $n$ elements are further randomly sampled from
$\boldsymbol{\mathcal{Z}}^{l+1}$. Therefore a set with
$\widetilde{m}=m \times n$ samples~(namely, there are $m$ images, and
each image provides $n$ elements for $L_{l+1}$) is obtained. We
introduce a new index $t$ to traverse those $\widetilde{m}$ elements,
and denote the corresponding subset of \textit{filters} in
$\boldsymbol{\mathcal{W}}^{l+1}$ as
$\boldsymbol{\mathcal{W}}^{l+1,t} \in \mathbb{R}^{J^{l} \times W^{l+1}
  \times W^{l+1}},\ 1 \le t \le \widetilde{m}$, and \textit{receptive
  fields}~\footnote{The region of the input space that is relative to
  the unit we are taking into consideration.} in
$\boldsymbol{\widehat{\mathcal{Z}}}^{l}$ as
$\boldsymbol{\widehat{\mathcal{R}}}^{l,t} \in \mathbb{R}^{J^{l} \times
  W^{l+1} \times W^{l+1}},\ 1 \le t \le \widetilde{m}$. What proposed
in ThiNet is actual a greedy solution of the following optimization
problem:
\begin{multline}
  \label{eq:thinet_optimization_problem}
  \qquad\argmin_{T}\sum_{t=1}^{\widetilde{m}}\left( \sum_{j \in
      T}\boldsymbol{\widehat{\mathcal{R}}}^{l,t}_{j::}\ast\boldsymbol{\mathcal{W}}^{l+1,t}_{j::} \right)^2\\
  \textrm{s.t.}\quad |T|=J^{l}\cdot\gamma,\ T \subset \{ 1, 2, \cdots, J^{l}\}
\end{multline}
in which $|T|$ denotes the number of selected indexes in $T$, and
$\boldsymbol{\widehat{\mathcal{R}}}^{l,t}_{j::}$,
$\boldsymbol{\mathcal{W}}^{l+1,t}_{j::}$ denote the $j$-th slide of
$\boldsymbol{\widehat{\mathcal{R}}}^{l,t}$ and
$\boldsymbol{\mathcal{W}}^{l+1,t}$. Those
$\boldsymbol{\widehat{\mathcal{Z}}}^{l}_{j::}$ with indexes fall in
$T$ are categorized as ``weak channels'' and are pruned. The
corresponding before-activation slides
$\{\boldsymbol{\mathcal{Z}}^{l}_{j::}, j \in T\}$ are implicated, and
are pruned as well. Consequently, the corresponding filters
$\{\boldsymbol{\mathcal{W}}^{l}_{:j::}, j \in T\}$ which generate
$\{\boldsymbol{\mathcal{Z}}^{l}_{j::}, j \in T\}$ are also
pruned. Please note that as three dimensional sub-tensors, the pruning
of $\{\boldsymbol{\mathcal{W}}^{l}_{:j::}, j \in T\}$ results in
cutting down of numerous parameters and FLOPs~(FLoating-point
OPerations) in the corresponding convolutions.

Fig.~\ref{fig:eq1_eq2_illustration} provides a toy example. In this toy example, the convolutional layer $L_l$ to be
pruned contains a before-activation output tensor
$\boldsymbol{\mathcal{Z}}^{l}$ with slide number $J^{l}=5$. Therefore
$\boldsymbol{\widehat{\mathcal{Z}}}^{l}$, the corresponding
after-activation input tensor for the next layer $L_{l+1}$ contains
five slides: $\boldsymbol{\widehat{\mathcal{Z}}}^{l}_{1::}$,
$\boldsymbol{\widehat{\mathcal{Z}}}^{l}_{2::}$,
$\boldsymbol{\widehat{\mathcal{Z}}}^{l}_{3::}$,
$\boldsymbol{\widehat{\mathcal{Z}}}^{l}_{4::}$, and
$\boldsymbol{\widehat{\mathcal{Z}}}^{l}_{5::}$. The next layer
$L_{l+1}$ only contains $K^{l+1}=1$ output channel/slide,
corresponding to $\boldsymbol{\mathcal{Z}}^{l+1}_{1::}$. The
pre-defined pruning rate $\gamma$ is set to 40\%, which means that two
out of the five slides in $\boldsymbol{\widehat{\mathcal{Z}}}^{l}$
should be pruned. Only one point/element is randomly sampled from
$\boldsymbol{\mathcal{Z}}^{l+1}_{1::}$, the only one slide of
$L_{l+1}$ in this toy example. Therefore Eq.(2) of the manuscript can
be simplified to:
\begin{equation*}
  \qquad\argmin_{T}\left( \sum_{j \in
      T}\boldsymbol{\widehat{\mathcal{R}}}^{l,1}_{j::}\ast\boldsymbol{\mathcal{W}}^{l+1,1}_{j::} \right)^2,
  \textrm{s.t.}\quad |T|=2,\ T \subset \{ 1, 2, \cdots, 5\}
\end{equation*}
From the equation above we can see
$\boldsymbol{\mathcal{W}}^{l+1,1}_{j::}=\boldsymbol{\mathcal{W}}^{l+1}_{j1::},\ 1
\le j \le 5$, as also denoted in
Fig.~\ref{fig:eq1_eq2_illustration}. For every filter,
$\boldsymbol{\mathcal{W}}^{l+1}_{j1::},\ 1 \le j \le 5$, the corresponding
receptive field in $\boldsymbol{\widehat{\mathcal{Z}}}^{l}$ is
$\boldsymbol{\widehat{\mathcal{R}}}^{l,1}_{j::},\ 1 \le j \le 5$,
respectively, as marked by dashed boxes in
Fig.~\ref{fig:eq1_eq2_illustration}. Since the $\ell^2$-norm of
$\boldsymbol{\widehat{\mathcal{R}}}^{l,1}_{1::}\ast\boldsymbol{\mathcal{W}}^{l+1,1}_{1::}+\boldsymbol{\widehat{\mathcal{R}}}^{l,1}_{3::}\ast\boldsymbol{\mathcal{W}}^{l+1,1}_{3::}$
is minimal, they are decimated. As a result,
$\boldsymbol{\mathcal{W}}^{l+1,1}_{1::}$ and
$\boldsymbol{\mathcal{W}}^{l+1,1}_{3::}$ are pruned. Two slides,
$\boldsymbol{\widehat{\mathcal{Z}}}^{l}_{1::}$ and
$\boldsymbol{\widehat{\mathcal{Z}}}^{l}_{3::}$ which contain
$\boldsymbol{\widehat{\mathcal{R}}}^{l,1}_{1::}$ and
$\boldsymbol{\widehat{\mathcal{R}}}^{l,1}_{3::}$ are pruned as
well. Though not illustrated in Fig.~\ref{fig:eq1_eq2_illustration},
we should note that in $L_l$, $\boldsymbol{\mathcal{Z}}^{l}_{1::}$,
$\boldsymbol{\mathcal{Z}}^{l}_{3::}$,
$\boldsymbol{\mathcal{W}}^{l}_{:1::}$,
$\boldsymbol{\mathcal{W}}^{l}_{:3::}$ are also pruned accordingly.

On the contrary, $l_1$-norm based scheme~\cite{li_iclr_2017} only
considers the statistical properties of the filters themselves. For a
given convolutional layer $L_l$, its $K^{l}$ output channels are
traversed. The $l_1$ norms of the corresponding filters
$\{\boldsymbol{\mathcal{W}}^{l}_{:k::},\ 1 \le k \le K^{l}\}$ are
calculated and sorted from lowest to highest. For a pre-defined
pruning rate $\gamma$, the first $K^{l}\cdot\gamma$ filters in the
sorted list are pruned. Those output channels corresponding to the
pruned filters are removed as well.

\subsection{Rationale of our proposed CALPA-NET}
\label{sec:rationale}

\begin{figure*}[!t]
  \centering
  \includegraphics[width=0.8\textwidth,keepaspectratio]{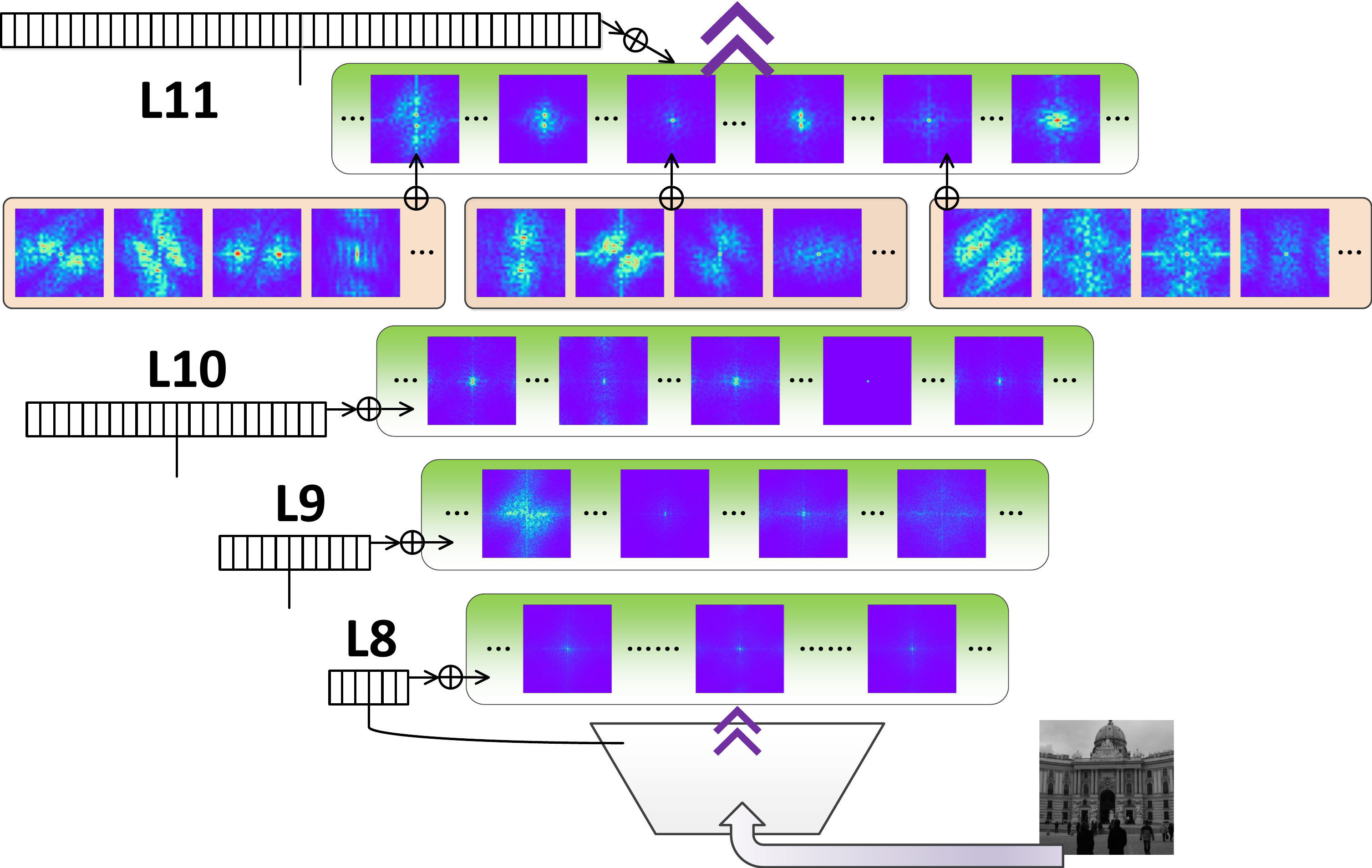}
  \caption[]{Amplitudes of the two-dimensional Fourier-transformed
    output channels~(visually represented with heat maps). The
    investigated target is a well-trained SRNet model aiming at detecting
    J-UNIWARD stego images with 0.4 bpnzAC payload and QF 75.  The
    input is BOSSBase image No.~1013.  Output channels of the second
    convolutional layer of ``Type 3'' residual blocks~(``L8'', ``L9'', ``L10'',
    and ``L11'') are investigated. For ``L11'', we further investigate the
    intermediate feature maps before the summation of the
    representative output channels.}
  \label{fig:psm_fft2d} 
\end{figure*} 

How to determine the width~(number of output channels/trainable
parameters) in each layer is a well-known open problem in deep
learning literature. As far as we know, in the field of deep-learning
based pattern recognition, there are no firm principles for
determination of layer width given a fixed amount of computational
resources. Therefore we opt to narrow our focus to deep-learning based
image steganalysis.

In the field of image steganalysis, a notable philosophy is
established since the reign of ``rich model+ensemble classifier''
solution~\cite{fridrich_tifs_2012}: model diversity is crucial to
success of steganalytic detectors. The adoption of ResNet-style
shortcut connections in latest deep-learning based steganalyzers
actually supports the above philosophy since ResNet-style shortcuts
resemble ensemble classifier, as pointed out in \cite{veit_nips_2016}.

In fact, in the field of deep-learning based pattern recognition,
upper-level representations~(usually output channels of upper-level
convolutional layers) learned with pre-trained deep CNNs are widely
used as classification feature
set~\cite{sharif_cvpr_2014,oquab_cvpr_2014}. Since deep CNN based
steganalysis is one of the subfields of deep-learning based pattern
recognition, it would be reasonable to believe that the argument which
makes sense in steganalytic feature-set construction can also be
established for the mid-level representations of deep-learning based
steganalyzers. Therefore, given a deep-learning based steganalyzer,
its upper-level convolutional layer $L_l$ can be regarded as a
features extractor. Its every output channel~(feature map) can be
viewed as a generated sub-model and should contains diverse
statistical patterns, according to \cite{fridrich_tifs_2012}.

From the perspective of time-frequency analysis, those ideal diverse
statistical patterns correspond to sparse intensity fluctuations in
nearly non-overlapping frequency subbands. Conversely, those strong
intensities in the most common frequency subbands~(usually those
low-frequency subbands) usually represent the underlying universal
patterns and contribute little to model diversity. However, as shown
below, the experimental evidence clearly show that the aggregation in
convolution with large enough input channels suppresses sparse
intensity fluctuations as well as heightens strong intensities in
low-frequency subbands. 

We take a well-trained SRNet model aiming at detecting J-UNIWARD stego
images with 0.4 bpnzAC payload and QF 75 for sample. With the
$256 \times 256$ JPEG representation of BOSSBase image No.~1013 as
input, we investigate the output channels of the second convolutional
layer of ``Type 3'' residual blocks~(``L8'', ``L9'', ``L10'', and
``L11''). \footnote{Please note that the output channels of the second
  convolutional layer of a given ``Type 3'' residual block, rather
  than the outputs of the entire residual block are investigated. This
  is due to the fact that with shortcut connection, the outputs of the
  entire ``Type 3'' residual block are the summation of the output
  channels of the second convolutional layer and the corresponding
  input channels of the entire residual block.} For the broader
``L11'', we further investigate the intermediate feature maps before
the summation of the representative output channels.

Fig.~\ref{fig:psm_fft2d} shows the heat maps of amplitudes of some
representative two-dimensional Fourier-transformed frequency-domain
output channels for those residual blocks~(please enlarge it for
better visual effect). From Fig.~\ref{fig:psm_fft2d} we can observe
that there are strong intensities in the most
common-frequency/low-frequency sub-bands of the outputs of the main
branch of the residual blocks. Even for those output channels with
drastic fluctuation, the evidence is obvious.

For ``L11'', three output channels of the second convolutional layer
are further selected as samples. The left one in Fig.~\ref{fig:psm_fft2d} is
with relatively obvious intensity fluctuations in high-frequency
subbands, the middle one only contains energy in low-frequency
sub-bands, while the right one is in somewhere between the two
extremes. For every one of the three output channels, four
intermediate feature maps before the summation are further
selected. All of those selected intermediate feature maps are with
obvious intensity fluctuations in high-frequency subbands. From
Fig.~\ref{fig:psm_fft2d} we can see that with large enough input
channels, for a given output channel, no matter how dispersive the
resulting intensity fluctuations in it, the summation/aggregation in
convolution of a given filter tensor do suppress the ever-present
sparse intensity fluctuations in higher-frequency subbands of the
intermediate feature maps generated by involved kernels.

Therefore we can make a straightforward inference that the broad
convolutional layers in the inverted-pyramid style upper modules of
existing deep-learning based steganalyzers might violate the model
diversity oriented philosophy. In Appendix.~\ref{app:reflection}, we
provide a theoretical reflection to support our argument.
\begin{figure}[!t]
  \centering
  \includegraphics[width=0.6\columnwidth,keepaspectratio]{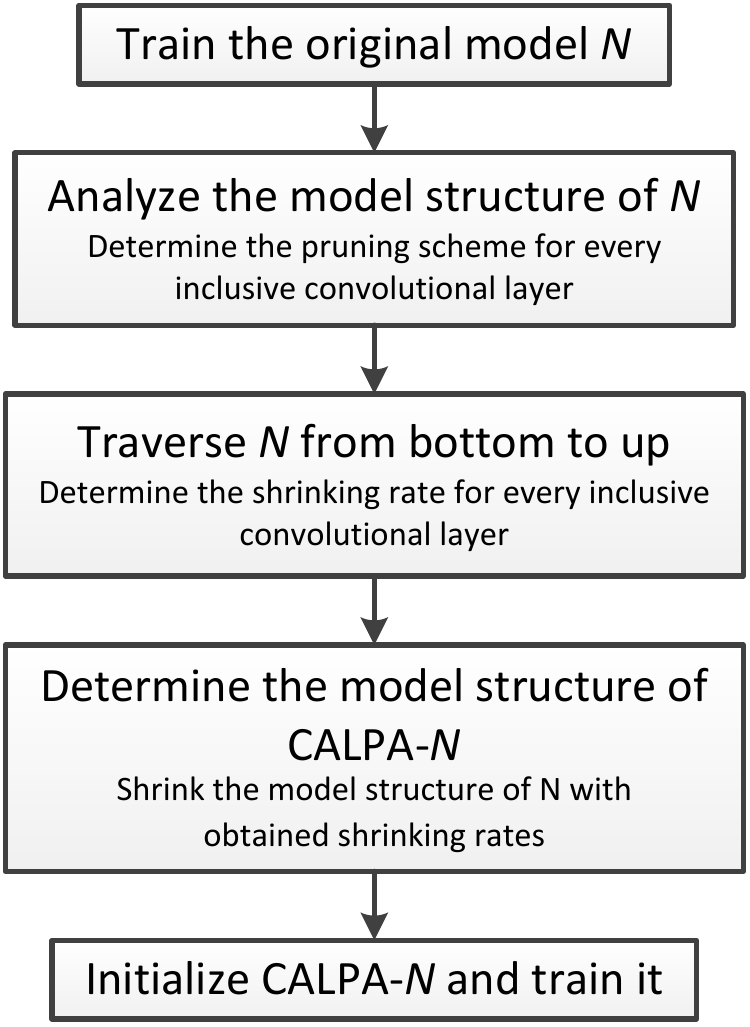} 
  \caption[]{The overall CALPA-NET diagram for a given deep-learning 
    based steganalytic model \textit{N}.}
  \label{fig:calpa_diagram}
\end{figure} 

\subsection{Detailed algorithm of our proposed CALPA-NET}
\label{sec:algorithm}

\subsubsection{The overall procedure}
\label{sec:overall_procedure}

Our proposed approach is named CALPA-NET, the ChAnneL-Pruning-Assisted
deep residual NETwork for image steganalysis. ``CALPA'' is homophonic
with ``KALPA'', a Sanskrit word for a never ending loop in which the
universe circularly expands and shrinks. The demonstration in
Sect.~\ref{sec:rationale} has revealed that the broad inverted-pyramid
structure might be a bloated solution for deep-learning based
steganalyzers. Therefore our target is to shrink the excessive
expanded structure of existing successful deep-learning based
steganalyzers to make them more cost-effective, as the word ``KALPA''
implies.

Recently, an interesting phenomenon that network pruning can be
regarded as one sort of network architecture searching approach is
reported in the field of computer vision~\cite{liu_iclr_2019}.  An
effective pruned network trained from scratch can show the performance
as good as the one with traditional three-stage
training-pruning-finetuning pipeline. This phenomenon inspires us to
use effective network pruning criteria to determine the most
cost-effective architecture for deep-learning based steganalyzers. The
diagram of the overall procedure is shown in
Fig.~\ref{fig:calpa_diagram}. Given a deep-learning based steganalytic
model \textit{N}, it is trained first using the original training
protocol. At the same time, the structure of \textit{N} is analyzed. A
specific pruning scheme is determined for every inclusive
convolutional layer in the pipeline according to its position in the
pipeline~(see Sect.~\ref{sec:criterion}). Then the well-trained
\textit{N} is traversed from bottom to top and the shrinking rate for
every inclusive convolutional layer is determined in a data-driven
manner. A brand new model structure is obtained via shrinking the
channels of every inclusive layer with corresponding shrinking
rate~(see Sect.~\ref{sec:architecture_search}). The resulting model is
referred to as CALPA-\textit{N}, and is reset and trained from
scratch. In this paper, two representative deep-learning based
steganalytic models are involved: SRNet~\cite{boroumand_tifs_2018} and
XuNet2~\cite{xu_ihmmsec_2017}. We call the corresponding CALPA-NET as
CALPA-SRNet and CALPA-XuNet2.

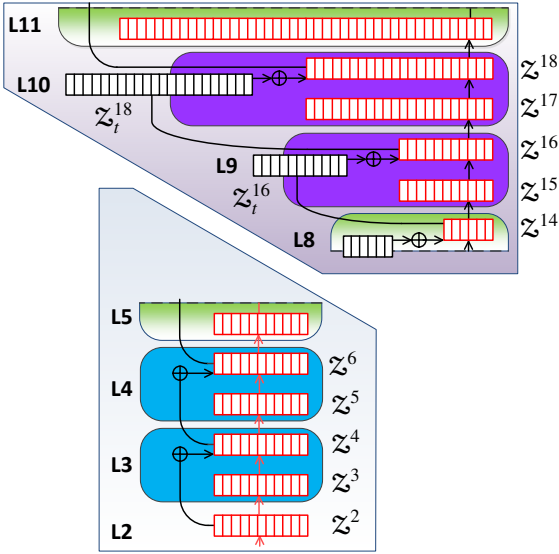
\begin{figure}[!t]
  \centering
  %\includegraphics[width=\columnwidth,keepaspectratio]{calpa_l1_thinet_analysis}  
  %\begin{overpic}[width=0.9\columnwidth,keepaspectratio,grid,tics=4]{calpa_l1_thinet_analysis}
  \begin{overpic}[width=0.9\columnwidth,keepaspectratio]{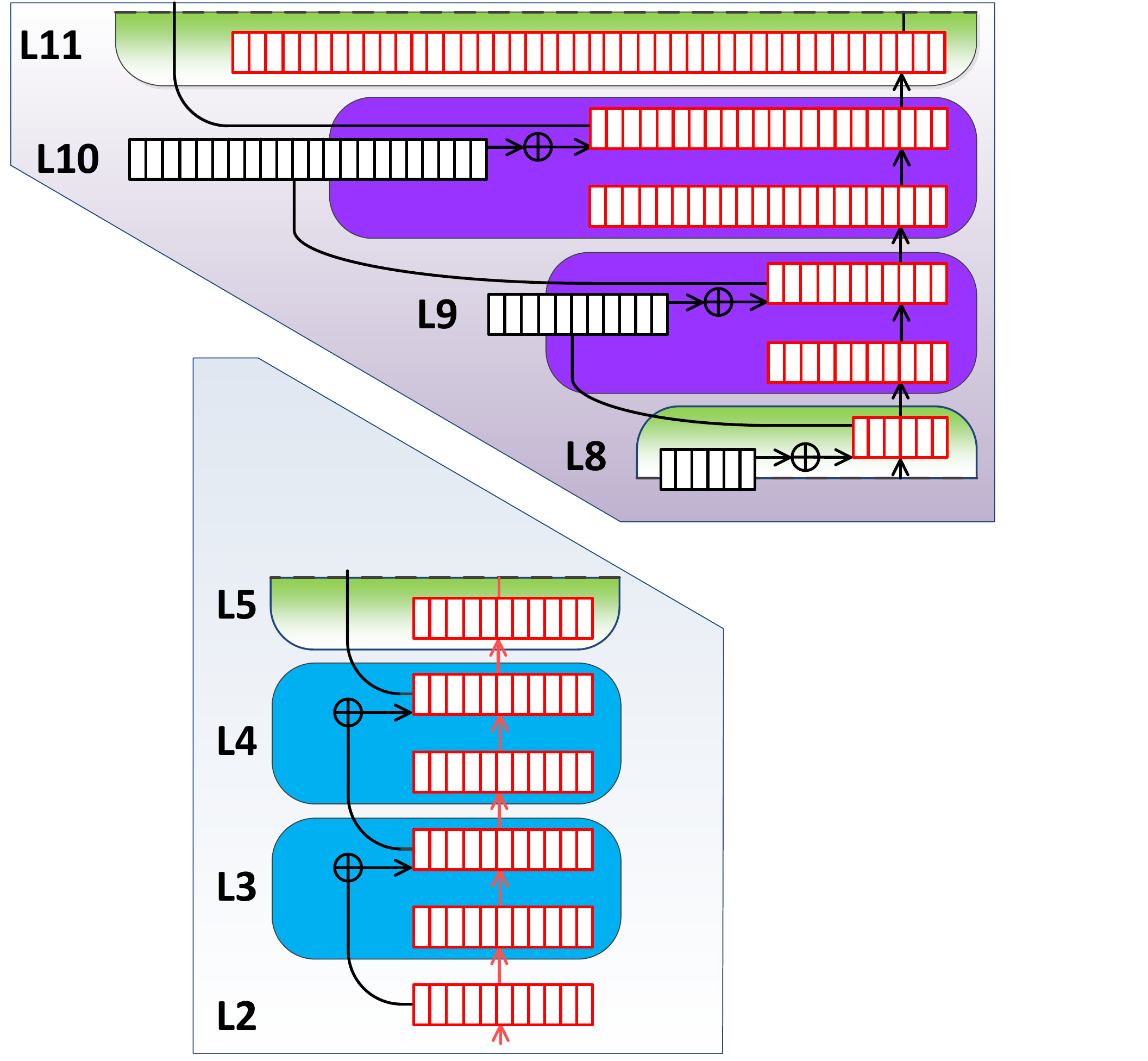}
    \put(55.5,3.5){$\boldsymbol{\mathcal{Z}}^{2}$}
    \put(55.5,10.5){$\boldsymbol{\mathcal{Z}}^{3}$}
    \put(55.5,17){$\boldsymbol{\mathcal{Z}}^{4}$}
    \put(55.5,23.5){$\boldsymbol{\mathcal{Z}}^{5}$}
    \put(55,30){$\boldsymbol{\mathcal{Z}}^{6}$}
    \put(87,53){$\boldsymbol{\mathcal{Z}}^{14}$}
    \put(87,59){$\boldsymbol{\mathcal{Z}}^{15}$}
    \put(87,66){$\boldsymbol{\mathcal{Z}}^{16}$}
    \put(39,58){$\boldsymbol{\mathcal{Z}}^{16}_{t}$}
    \put(87,73){$\boldsymbol{\mathcal{Z}}^{17}$}
    \put(87,79.5){$\boldsymbol{\mathcal{Z}}^{18}$}
    \put(16,71.5){$\boldsymbol{\mathcal{Z}}^{18}_{t}$}
  \end{overpic}
  \caption[]{The shortcut connections in two types of residual blocks
    in SRNet. ``$\oplus$" denotes element-wise addition.}
  \label{fig:calpa_l1_thinet_analysis}
\end{figure} 

\subsubsection{The proposed hybrid channel pruning criterion}
\label{sec:criterion}

Unfortunately, there is no known channel pruning schemes for shortcut
connections yet. Our preferred channel pruning criterion is the one
used in ThiNet~\cite{luo_iccv_2017} due to its simplicity and
efficiency. However, the ThiNet scheme cannot be used in shortcut
connections.

An example from SRNet is given in
Fig.~\ref{fig:calpa_l1_thinet_analysis}. Two types of residual blocks
are picked out from Fig.~\ref{fig:calpa_xunet_srnet_structure}. Blocks
``L3'' and ``L4'' with blue background contain direct shortcut
connections, while blocks ``L9'' and ``L10'' with pink background
contain transformed shortcut connections.  Please note that for direct
shortcut connections, the element-wise addition of the two relevant
convolutional layers demands that the sizes of the corresponding
output tensors, like $\boldsymbol{\mathcal{Z}}^{2}$ and
$\boldsymbol{\mathcal{Z}}^{4}$, $\boldsymbol{\mathcal{Z}}^{4}$ and
$\boldsymbol{\mathcal{Z}}^{6}$, should be the same. Further more, all
of the relevant convolutional layers linked via shortcut connections
should be with the same output tensor size. From
Fig.~\ref{fig:calpa_l1_thinet_analysis} it is clear that the size of
$\boldsymbol{\mathcal{Z}}^{2}$, $\boldsymbol{\mathcal{Z}}^{4}$ and
$\boldsymbol{\mathcal{Z}}^{6}$ must be the same. For transformed
shortcut connections, each shortcut is equipped with a specific
convolutional layer consisting of $1 \times 1$ kernels to extend the
size of the shortcut output tensor. Given a main-branch output tensor
$\boldsymbol{\mathcal{Z}}^{l}$, denote the output tensor in the
transformed shortcut connection element-wisely added to it as
$\boldsymbol{\mathcal{Z}}^{l}_{t}$. Analogously, the two convolutional
layers relevant to the element-wise addition in transformed shortcut
connections, like $\boldsymbol{\mathcal{Z}}^{16}$ and
$\boldsymbol{\mathcal{Z}}^{16}_{t}$, $\boldsymbol{\mathcal{Z}}^{18}$
and $\boldsymbol{\mathcal{Z}}^{18}_{t}$, should also be the same.

In ThiNet, the pruning in $\boldsymbol{\mathcal{Z}}^{l}$ relies on the
samples in the output tensor on top of it. However, the introduction
of shortcut connections implies that $\boldsymbol{\mathcal{Z}}^{l}$
may correspond to two or even more tensors on top of it. As shown in
Fig.~\ref{fig:calpa_l1_thinet_analysis}, for direct shortcut
connections, the pruning in $\boldsymbol{\mathcal{Z}}^{2}$ relies on
$\boldsymbol{\mathcal{Z}}^{3}$ and $\boldsymbol{\mathcal{Z}}^{4}$, and
$\boldsymbol{\mathcal{Z}}^{4}$ in turns replies on
$\boldsymbol{\mathcal{Z}}^{5}$ and $\boldsymbol{\mathcal{Z}}^{6}$. for
transformed shortcut connections, both $\boldsymbol{\mathcal{Z}}^{16}$
and $\boldsymbol{\mathcal{Z}}^{16}_{t}$ rely on
$\boldsymbol{\mathcal{Z}}^{17}$ and
$\boldsymbol{\mathcal{Z}}^{18}_{t}$. Therefore, ThiNet is not suitable
for shortcut connections. As a result, a hybrid channel pruning
criterion is employed in our proposed approach:

For those convolutional layers not involved in shortcut connections,
ThiNet scheme is applied to them to determine the shrinking rates of
their output tensors. As demonstrated in
Fig.~\ref{fig:calpa_l1_thinet_analysis},
$\boldsymbol{\mathcal{Z}}^{3}$, $\boldsymbol{\mathcal{Z}}^{5}$,
$\boldsymbol{\mathcal{Z}}^{15}$ and $\boldsymbol{\mathcal{Z}}^{17}$
are not involved in shortcut connections. ThinNet scheme is applied to
them and their shrinking rates are determined according to the output
tensors on top of them, $\boldsymbol{\mathcal{Z}}^{4}$,
$\boldsymbol{\mathcal{Z}}^{6}$, $\boldsymbol{\mathcal{Z}}^{16}$ and
$\boldsymbol{\mathcal{Z}}^{18}$, respectively.

For those relevant convolutional layers linked via direct shortcut
connections, they are handled altogether with a post-processing
mechanism. $l_1$-norm based scheme~\cite{li_iclr_2017} is applied to
the output tensor of the lowest convolutional layer and determine its
shrinking rate. The same shrinking rate is assigned to the output
tensor of the rest relevant layers. Take SRNet for example,
$\boldsymbol{\mathcal{Z}}^{2}$, $\boldsymbol{\mathcal{Z}}^{4}$,
$\boldsymbol{\mathcal{Z}}^{6}$, and even
$\boldsymbol{\mathcal{Z}}^{8}$, $\boldsymbol{\mathcal{Z}}^{10}$,
$\boldsymbol{\mathcal{Z}}^{12}$~(do not appear in
Fig.~\ref{fig:calpa_l1_thinet_analysis}) are linked via direct
shortcut connections. $l_1$-norm based scheme is applied to
$\boldsymbol{\mathcal{Z}}^{2}$, the lowest one. Then the obtained
shrinking rate is assigned to all the linked output tensors.

For those transformed shortcut connections, the two convolutional
layers relevant to the element-wise addition in transformed shortcut
connections are handled together. Given a main-branch output tensor
$\boldsymbol{\mathcal{Z}}^{l}$, its shrinking rate is determined via
$l_1$-norm based scheme. The same shrinking rate is assigned to
$\boldsymbol{\mathcal{Z}}^{l}_{t}$, the output tensor in the
corresponding transformed shortcut connection as well. Also take SRNet
for example, $l_1$-norm based scheme is used to determine the shrinking
rate of $\boldsymbol{\mathcal{Z}}^{16}$ and
$\boldsymbol{\mathcal{Z}}^{18}$, and then the obtained shrinking rate is
assigned to $\boldsymbol{\mathcal{Z}}^{16}_{t}$ and
$\boldsymbol{\mathcal{Z}}^{18}_{t}$, respectively.
\subsubsection{Channel-pruning-assisted architecture search}
\label{sec:architecture_search}

\begin{algorithm}
  \caption{Shrinking rate determination algorithm for $L_l$ and all the
    relevant convolutional layers.}
  \label{alg:shrinking_rate_determination}
  \begin{algorithmic}[1]
    \Require A well-trained model \textit{N}, a standalone validation
    dataset $D_v$, a pre-defined step $\epsilon$~(set to $5\%$ in our
    experiments), and a tolerable accuracy lost $\varsigma$~(set to
    $5\%$ in our experiments).
    \State Initialize the pruning rate $\gamma_l=0\%$.
    \State \algmultiline{%
      Validate \textit{N} in $D_v$. Denote the obtained validation
      accuracy as $Acc_0$. Let $Acc_{p_1}=Acc_0$.}
    \Loop
    \State Set $\gamma_l=\gamma_l+\epsilon$. \label{renew_gamma}
%     \algstore{shrinking_rate_determination}
%   \end{algorithmic}
% \end{algorithm}
% \begin{algorithm}
%   \begin{algorithmic}[1]
%     \algrestore{shrinking_rate_determination}
    \If {$L_l$ is not in shortcut connections}
    \State use ThiNet to prune $L_l$.
    \Else
    \State \algmultiline{%
      use $l_1$-norm based scheme to prune $L_l$ and all the relevant
      convolutional layers linked with it via shortcut connections.}
    \EndIf
    \State \algmultiline{%
      Let $N_p$ denotes the pruned model. Validate $N_p$ in $D_v$ and
      denote the validation accuracy as $Acc_{p_2}$.}
    \If {$Acc_{p_1}-Acc_{p_2}>\varsigma$}
    \State \algmultiline{%
      an obvious accuracy decline is observed.}
    \Break
    \ElsIf {$Acc_0-Acc_{p_2}>\varsigma$}
    \State \algmultiline{%
      the detection accuracy has gradually declined to beyond the
      tolerance threshold.}
    \Break
    \EndIf
    \State Let $Acc_{p_1}=Acc_{p_2}$.
    \EndLoop
    \State \algmultiline{%
      set $\gamma_l=\gamma_l-\epsilon$, namely roll back to prior
      pruning rate.}
    \State \algmultiline{%
      Set the corresponding shrinking rate $\zeta_l=100\%-\gamma_l$. If
      $L_l$ is in shortcut connections, the shrinking rates of all the
      relevant convolutional layers linked with it via shortcut
      connections are set to $\zeta_l$ as well.}
  \end{algorithmic}
\end{algorithm}

Please note that in CALPA-NET, channel pruning approaches is used to
assist the determination of the shrinking rate of every involved
convolutional layer. A new term ``shrinking rate'' is introduced in
order to highlight the difference between our proposed CALPA-NET and
existing channel pruning approaches. Given a convolutional layer $L_l$
with a determined shrinking rate $\zeta_l$, the corresponding pruning
rate used in channel pruning approaches is $\gamma_l=1-\zeta_l$.

As mentioned in Sect.~\ref{sec:overall_procedure}, the well-trained
model \textit{N} is traversed from bottom to top for every inclusive
convolutional layer $L_l$ with shrinking rate undetermined. The
detailed algorithm is shown in
Algorithm~\ref{alg:shrinking_rate_determination}.

After the bottom-up traversal, every inclusive convolutional layer
$L_l$ is assigned a determined shrinking rate $\zeta_l$. The resulting
CALPA-\textit{N} model is obtained via shrinking the volume of
$\boldsymbol{\mathcal{Z}}^{l}$, the before-activation output tensor of
every inclusive $L_l$ to
$\mathbb{R}^{K^{l}_{\zeta} \times H^{l} \times W^{l}}$, in which
$K^{l}_{\zeta}=K^{l}\cdot\zeta_l$. The obtained CALPA-\textit{N} is
reset and trained from scratch.

\section{Experiments}
\label{sec:exp}

\subsection{Experiment setup}
\label{sec:setup}

\begin{table*}[!t]
  \centering
  \caption[]{Effect of different tolerable accuracy lost $\varsigma$
    on the parameters, FLOPs, and detection accuracies of
    CALPA-SRNet. The detection accuracies were obtained on the
    stand-alone testing set. The Two training scenarios are compared:
    ``trained from scratch'' and ``finetuned''. The trained models are
    aiming at detecting J-UNIWARD stego images with 0.4 bpnzAC payload
    and QF 75. The percentages of parameters and FLOPs of CALPA-SRNet
    compared to original SRNet are listed in parentheses. For
    accuracies, the terms in parentheses with preceding
    \textdownarrow\ and \textuparrow\ denote decrement or increment
    compared to original SRNet.}
  \resizebox{0.95\textwidth}{!}{%
    {\renewcommand{\arraystretch}{2}% for the vertical padding
      \begin{tabular}{ccccccc}
        \Xhline{2\arrayrulewidth}
        \multirow{2}{*}{} & \multirow{2}{*}{\large \makecell{\textbf{Original} \\
        \textbf{SRNet}}} & 
                           \multicolumn{5}{ c }{\large \textbf{CALPA-SRNet with a given
                           tolerable accuracy lost $\varsigma$}}\\
        \cline{3-7}
                          & & 2\% & 5\% & 10\% & 20\% & 50\% \\
        \hline
        \textbf{Parameters} & 477.06$\times 10^4$ & 8.43$\times 10^4$~\textbf{(1.77\%)}& 6.93$\times 10^4$~\textbf{(1.45\%)}& 4.71$\times 10^4$~\textbf{(0.99\%)}& 3.45$\times 10^4$~\textbf{(0.72\%)}& 2.24$\times 10^4$~\textbf{(0.47\%)}\\
        \textbf{FLOPs} & 5.95$\times 10^9$ & 2.29$\times 10^9$~\textbf{(38.5\%)}& 1.97$\times 10^9$~\textbf{(33.1\%)}& 1.54$\times 10^9$~\textbf{(25.9\%)}& 1.37$\times 10^9$~\textbf{(23.1\%)}& 0.75$\times 10^9$~\textbf{(12.7\%)}\\
        \multirow{4}{*}{\textbf{Accuracy}} & \multirow{4}{*}{92.98\%} & \multicolumn{5}{ c }{\textbf{Trained from scratch}}\\
        \cline{3-7}                                                                      
                          & & 93.12\%~\textbf{(\textuparrow 0.14\%)}& 92.86\%~\textbf{(\textdownarrow 0.12\%)}& 92.63\%~\textbf{(\textdownarrow 0.35\%)}& 92.36\%~\textbf{(\textdownarrow 0.62\%)}& 87.94\%~\textbf{(\textdownarrow 5.04\%)}\\
                          & & \multicolumn{5}{ c }{\textbf{Finetuned}}\\
        \cline{3-7}                                                 
                          & & 93.07\%~\textbf{(\textuparrow 0.09\%)} & 92.8\%~\textbf{(\textdownarrow 0.18\%)}& 92.84\%~\textbf{(\textdownarrow 0.14\%)} & 92.59\%~\textbf{(\textdownarrow 0.39\%)} & 87.63\%~\textbf{(\textdownarrow 5.35\%)} \\
        \Xhline{2\arrayrulewidth}
      \end{tabular} 
    }} 
  \label{tab:threshold_candidates}
\end{table*}

The primary image dataset used in our experiments is the union of
BOSSBase v1.01~\cite{bas_ih_2011} and
BOWS2~\cite{bows2_link}\footnote{For a fair comparison, we tried our
  best to adopt almost the same experiment setup as which
  SRNet~\cite{boroumand_tifs_2018} was evaluated on.}, each of which
contains 10,000 $512\times512$ grayscale spatial images. All of the
images were resized to $256\times256$ using
Matlab$^{\textrm{\textregistered}}$ function \textit{imresize}. The
corresponding JPEG images were further generated with QFs~(Quality
Factors) 75 and 95. In every experiments, 10,000 BOWS2 images and
4,000 randomly selected BOSSBase images were used for
training. Another 1,000 randomly selected BOSSBase images were for
validation. The remaining 5,000 BOSSBase images were retained for
testing. Furthermore, following the generation pipeline mentioned in
\cite{boroumand_tifs_2018}, a subset of ImageNet
CLS-LOC~\cite{russakovsky_ijcv_2015} dataset with 250,000 grayscale
$256 \times 256$ JPEG images was also introduced to demonstrate the
performance of CALPA-NET on large-scale dataset under cover source
diversity scenario. ALASKA~\cite{cogranne_ihmmsec_2019} is another
large-scale dataset introduced to evaluate the performance of
CALPA-NET under stego source diversity scenario. Its JPEG
compressed~(with QF 75) grayscale $256\times 256$ version~(v2) was
adopted~\footnote{\url{http://alaska.utt.fr/ALASKA_v2_JPG_256_QF75_GrayScale.sh}}.

Four representative steganographic schemes, UERD~\cite{guo_tifs_2015}
and J-UNIWARD~\cite{holub_eurasip_2014} for JPEG domain, and
HILL~\cite{li_icip_2014} and S-UNIWARD~\cite{holub_eurasip_2014} for
spatial domain, were our attacking targets in the experiments. For
JPEG steganographic algorithms, the embedding payloads were set to 0.2
and 0.4 bpnzAC~(bits per non-zero AC DCT coefficient). For spatial
domain steganographic algorithms, the embedding payloads were set to
0.2 and 0.4 bpp~(bits per pixel).

XuNet2~\cite{xu_ihmmsec_2017} and SRNet~\cite{boroumand_tifs_2018}
were selected as the two initial architectures of our proposed
CALPA-NET. Our implementation of CALPA-NET and its corresponding
initial architectures are based on
TensorFlow~\cite{tensorflow2015-whitepaper}. Unless otherwise
specified, the two initial architectures were trained with the
hyper-parameters mentioned in the corresponding original papers. The
batch size in the training procedure was set to 32~(namely 16
cover-stego pairs). The maximum number of iterations was set to
$50 \times 10^4$ for SRNet, and $32 \times 10^4$ for
XuNet2. CALPA-XuNet2 and CALPA-SRNet adopted the same maximum number
of iterations as their corresponding initial architectures. Like their
corresponding initial architectures, the optimizers and learning rate
schedules used for CALPA-SRNet and CALPA-XuNet2 were specifically
defined as follows: CALPA-SRNet were trained using Adamax with
learning rate starting from 0.001. After $40 \times 10^4$ iterations
the learning rate was reduced to 0.0001. CALPA-XuNet2 was trained
using mini-batch stochastic gradient descent with learning rate
starting from 0.001. The learning rate was set to decrease $10\%$ for
every 5,000 iterations. The momentum was fixed to 0.9.

In each experiment, unless otherwise specified, the model was
validated and saved every one epoch~(in primary dataset,
$14,000/16=875$ iterations). The one with the best validation accuracy
was evaluated on the corresponding testing set. All of the experiments
were conducted on a GPU cluster with sixteen
NVIDIA$^{\textrm{\textregistered}}$ Tesla$^{\textrm{\textregistered}}$
P100 GPU cards. Bounded by computational resources, every experiment
was repeated three times, and the mean of the results on testing set
were reported. The source codes and auxiliary materials are available
for download from
GitHub~\footnote{\url{https://github.com/tansq/CALPA-NET}}.

\subsection{Compactness of CALPA-NET}
\label{sec:analysis_of_calpa_net}

In this section, we take CALPA-SRNet as example to analyze its
compactness and effectiveness.

\subsubsection{Compactness of CALPA-NET with different $\varsigma$}
\label{sec:analysis_of_calpa_net_1}

In Tab.~\ref{tab:threshold_candidates}, we compare the effect of
different tolerable accuracy lost $\varsigma$ on the model parameters,
FLOPs, and detection accuracies of CALPA-SRNet.

During calculation, the number of parameters and the number of FLOPs
for every convolutional layer are determined as follows~(let
$|\bullet|$ denotes the number of the corresponding variable):
$|$parameters$|$=$|$input
channels$|$$\times$$|$kernel
size$|$$\times$$|$output channels$|$; $|$FLOPs$|$=$|$output channel
size$|$$\times$$|$parameters$|$.

Here our adopted ``trained from scratch'' scenario is compared with
the traditional ``training-pruning-finetuning'' pipeline. In
``training-pruning-finetuning'' pipeline, the weights in non-pruned
kernels of the model were kept. Then the pruned model was
re-trained/finetuned for another $50 \times 10^4$ iterations. From
Tab.~\ref{tab:threshold_candidates} we can see that with the rise of
tolerable accuracy lost $\varsigma$, the parameters and FLOPs required
by CALPA-SRNet significantly decrease. However, with a mild
$\varsigma$~($2\%\sim 20\%$), the detection accuracy of the shrunken
model is close to the original SRNet. Another notable thing is that
the detection performances achieved by the shrunken models trained
from scratch are almost equivalent to those with
``training-pruning-finetuning'' pipeline. Such a phenomenon indicates
that in CALPA-Net, the efficient network architecture obtained via
channel-pruning-assisted search is critical.

By contrasting the benefits of the reduction of parameters and FLOPs
with the losses of detection performance, $\varsigma$ is set to $5\%$
in our final proposed framework. Please note that when
$\varsigma=5\%$, CALPA-SRNet achieves similar detection performance as
the original SRNet with mere $1.45\%$ parameters~(6.93$\times 10^4$
vs. 477.06$\times 10^4$) and $33.1\%$ FLOPs~(1.97$\times 10^9$
vs. 5.95$\times 10^9$). The corresponding architecture of CALPA-SRNet
can be referred back to
Fig.~\ref{fig:calpa_xunet_srnet_structure}. From
Fig.~\ref{fig:calpa_xunet_srnet_structure} we can see CALPA-SRNet
presents a slender bottleneck-like structure in contrast to
traditional deep-learning based steganalyzers.

However, please keep in mind that besides the complexity of the
deep-learning model itself, there are quite a few other factors which
influence the training efficiency, including access speed of the hard
disk where the data is stored in, size of the hard disk cache/memory
cache, speed of the high-bandwidth bridge on the motherboard. In order
to maximize the benefits offered by CALPA-SRNet, high-end hardwares
such as fast SSDs~(solid-state drives) are highly
recommended. Furthermore, please note that during the training
procedure, the overwhelming majority of GPU memory allocation is used
to store the input image batch and the corresponding input/output
channels of all layers, rather than the model parameters. Therefore
the superiority of CALPA-NET is its significantly reduced parameters
and FLOPs, not smaller GPU memory allocation.

\subsubsection{Tolerance to pruning rate in our approach}
\label{sec:analysis_of_calpa_net_2}

\begin{figure*}[!t]
  \centering
  \subfloat[]{
    \label{fig:thinet_sensitivity}
  \includegraphics[width=0.8\linewidth,keepaspectratio]{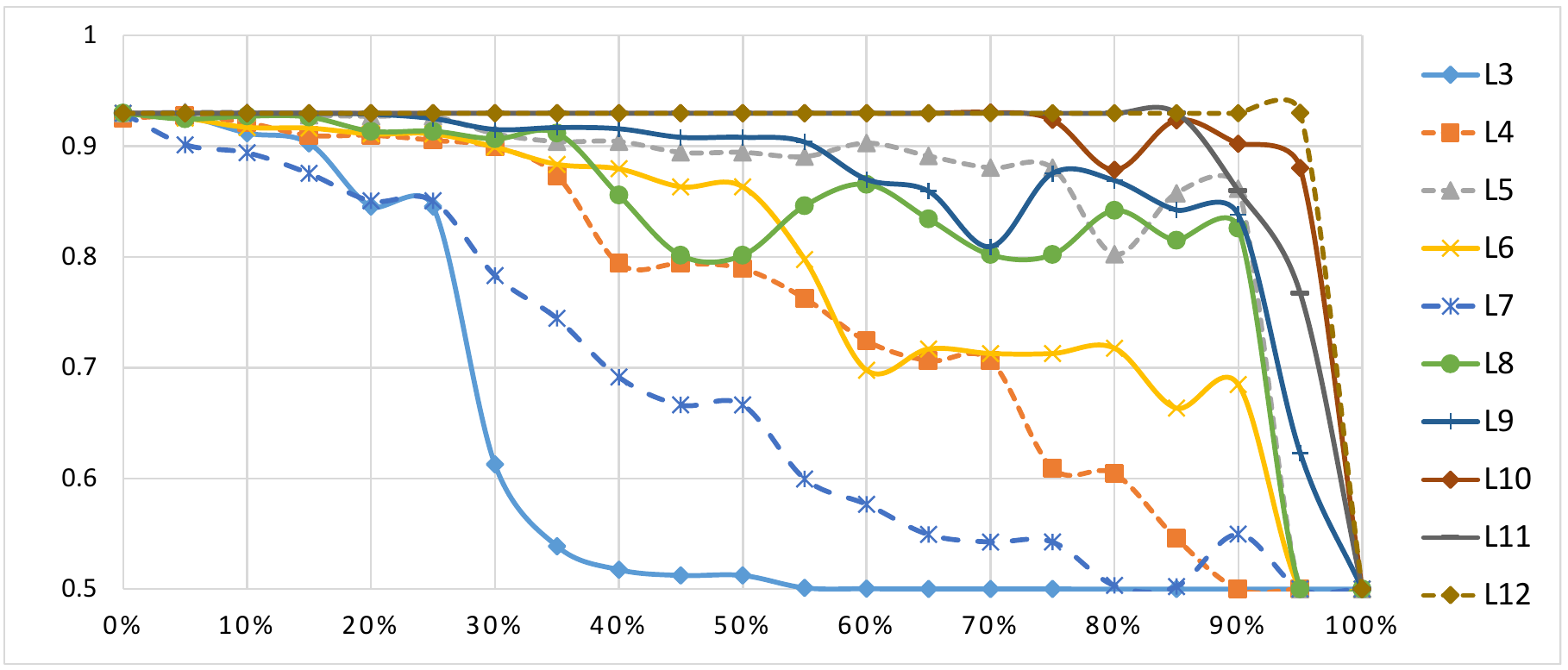}
  }\\
  \subfloat[]{
    \label{fig:li_sensitivity}
    \includegraphics[width=0.8\linewidth,keepaspectratio]{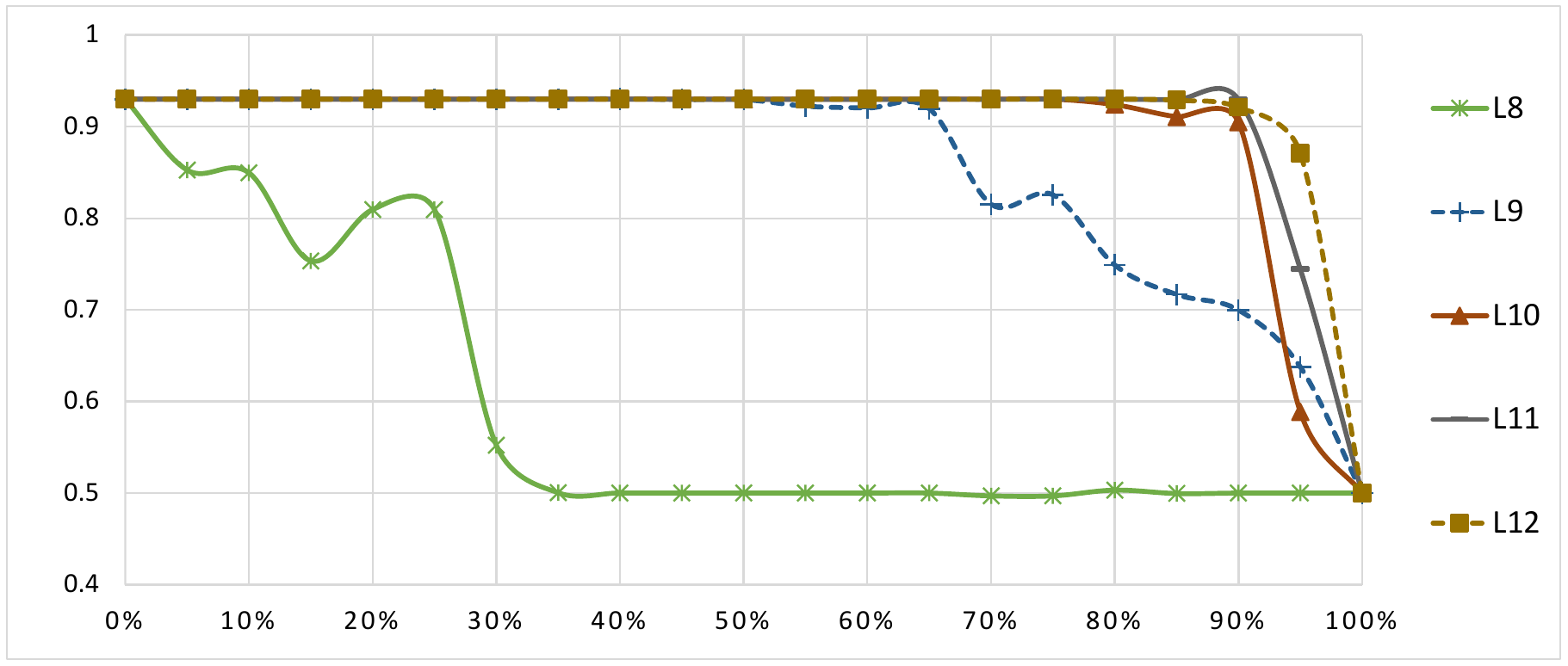}
  }\\
  \caption[]{Validation accuracies vs. growing pruning rates when
    determining the shrinking rate of every inclusive convolutional
    layer of CALPA-SRNet. The trained models are aiming at detecting J-UNIWARD stego images with 0.4
    bpnzAC payload and QF
    75. $\varsigma=5$. \subref{fig:thinet_sensitivity} is for ThiNet
    scheme in blocks ``L3''---``L12''.  \subref{fig:li_sensitivity} is
    for $l_1$-norm based scheme in transformed shortcut connections of
    blocks ``L8''---``L12''.}
  \label{fig:thinet_li_sensitivity}
\end{figure*}

In Fig.~\ref{fig:thinet_li_sensitivity}, we show how the validation
accuracy changes with successive growing pruning rates when
determining the shrinking rate of every inclusive convolutional layer
of CALPA-SRNet. In Algorithm~\ref{alg:shrinking_rate_determination},
we just run the loop~(from Line 3 to Line 19) for $\gamma_l$ from 0\%
till 100\%, and disable the early exit conditions from Line 11 to Line
17 to obtain the corresponding validation accuracies. From
Fig.~\ref{fig:thinet_li_sensitivity} no particular patterns can be
observed. However, no matter for ThiNet scheme~(as shown in
Fig.~\ref{fig:thinet_li_sensitivity}\subref{fig:thinet_sensitivity}),
or for $l_1$-norm based scheme~(as shown in
Fig.~\ref{fig:thinet_li_sensitivity}\subref{fig:li_sensitivity}),
convolutional layers in top blocks always present high tolerance to
severe pruning rates.

\subsubsection{Effectiveness of CALPA-NET compared to traditional
  ``training-pruning-finetuning'' pipeline}
\label{sec:analysis_of_calpa_net_3}

\begin{figure*}[!t]
  \centering
  \includegraphics[width=\textwidth,keepaspectratio]{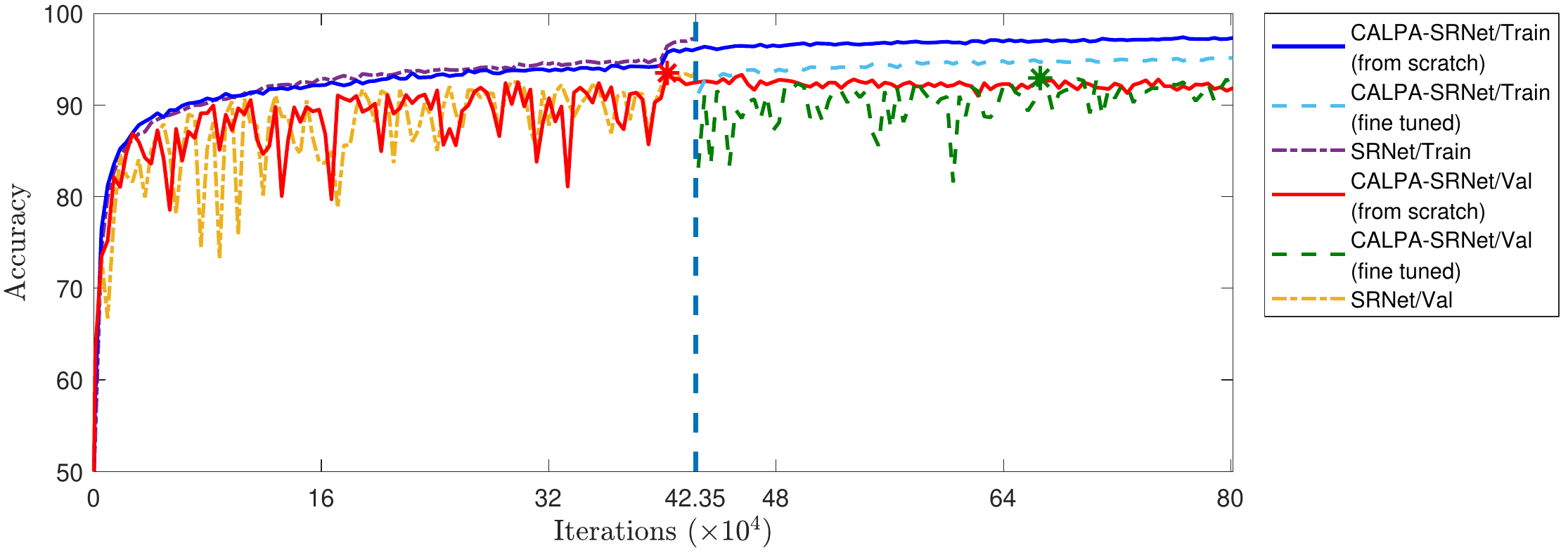}  
  \caption[]{Comparison of the training accuracies, validation
    accuracies and testing accuracies vs. training iterations for the
    original SRNet, CALPA-SRNet trained from scratch, and CALPA-SRNet
    with ``training-pruning-finetuning'' pipeline. The trained models
    are aiming at detecting J-UNIWARD stego images with 0.4 bpnzAC
    payload and QF 75. $\varsigma=5$. ``\ding{107}'' denotes the
    point with the highest value at the corresponding plot. }
  \label{fig:comp_fine_tuned_from_scratch}
  \vspace{2em}
\end{figure*} 

\begin{figure*}[!t]
  \centering
  \subfloat[]{
    \label{fig:srnet_jpg75_juniward_0204}
    \includegraphics[width=0.4\textwidth,keepaspectratio]{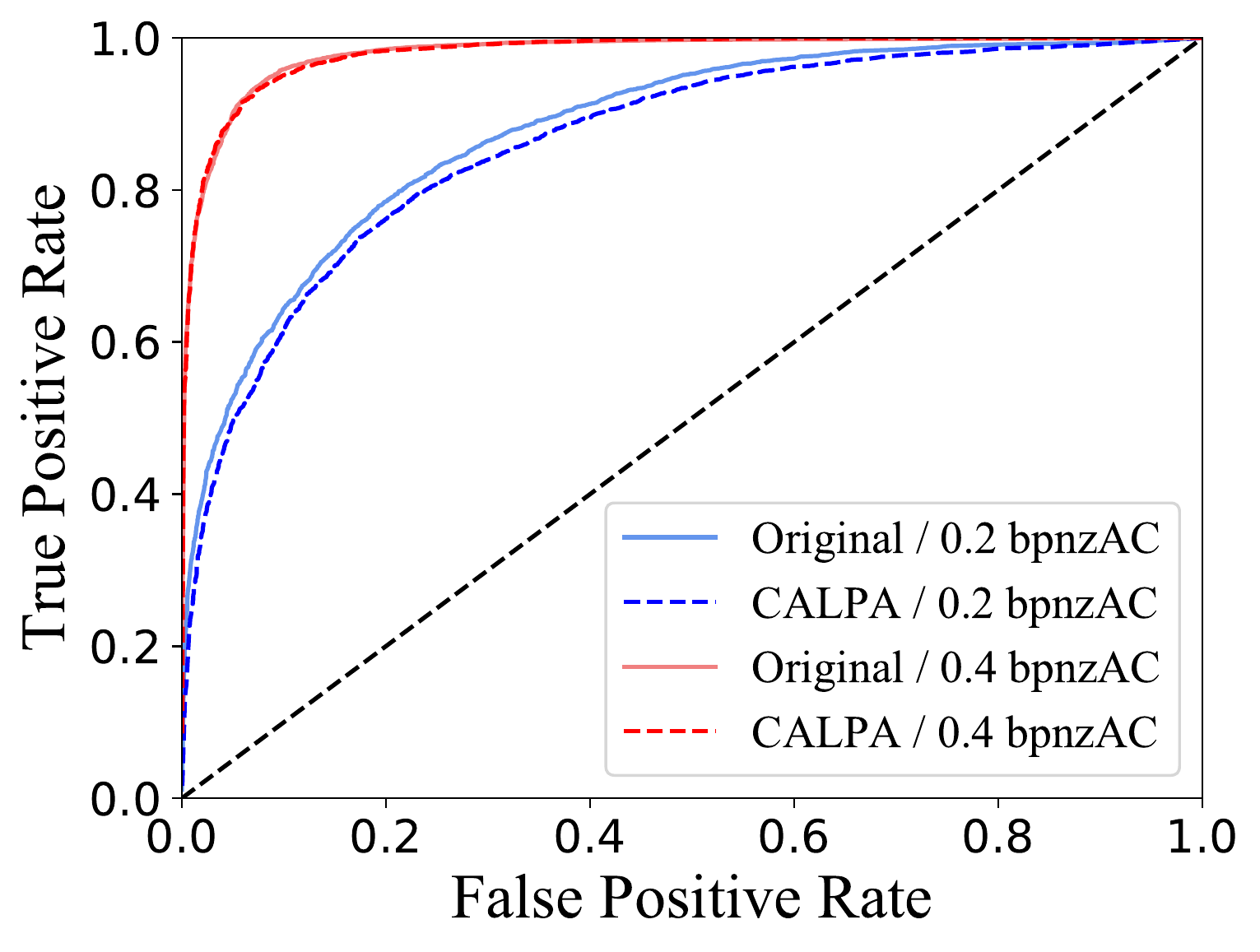} 
  }\hspace{.05\linewidth}
  \subfloat[]{
    \label{fig:srnet_jpg75_uerd_0204}
    \includegraphics[width=0.4\textwidth,keepaspectratio]{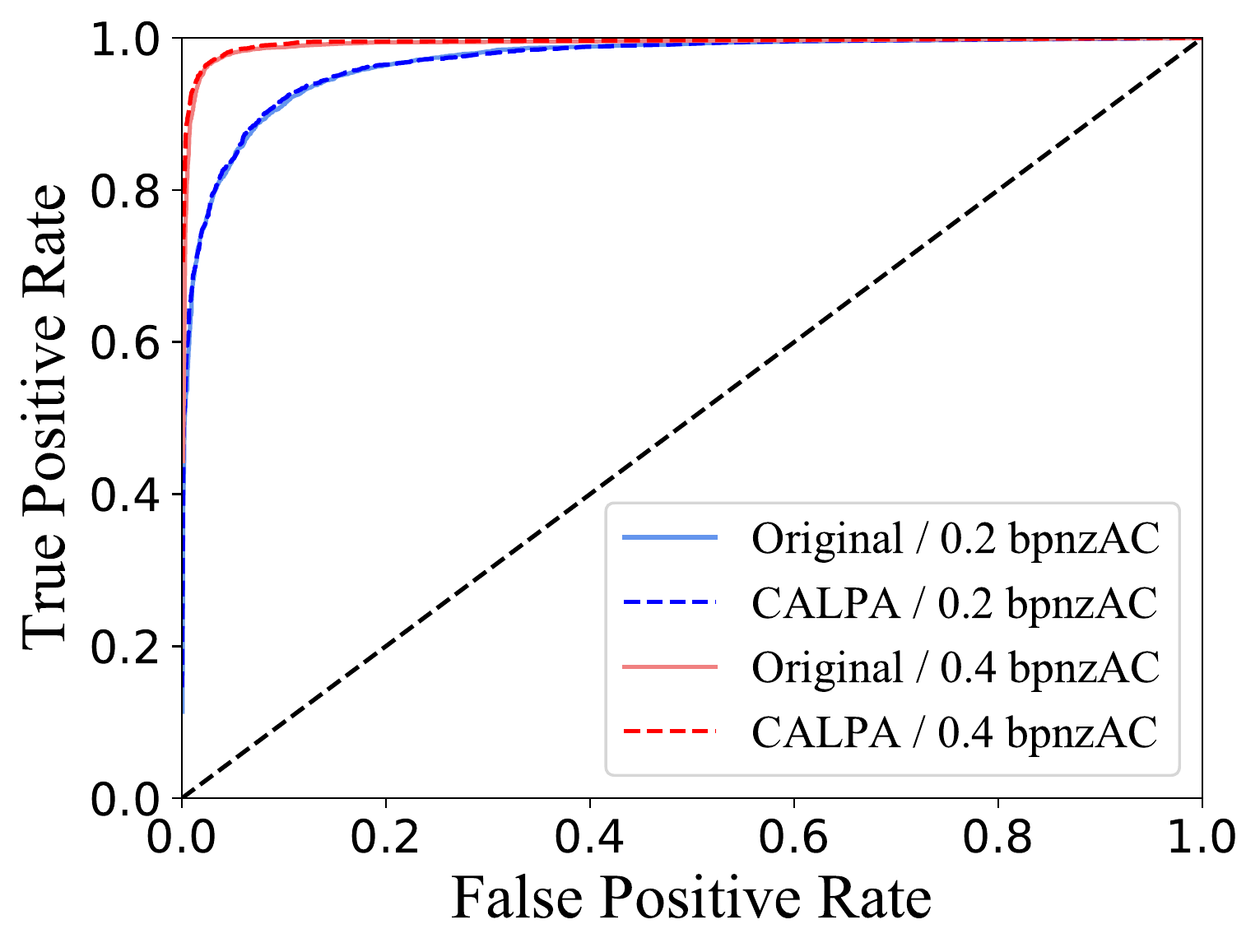} 
  }\\
  \subfloat[]{
    \label{fig:srnet_jpg95_uerd_juniward_04}
    \includegraphics[width=0.4\textwidth,keepaspectratio]{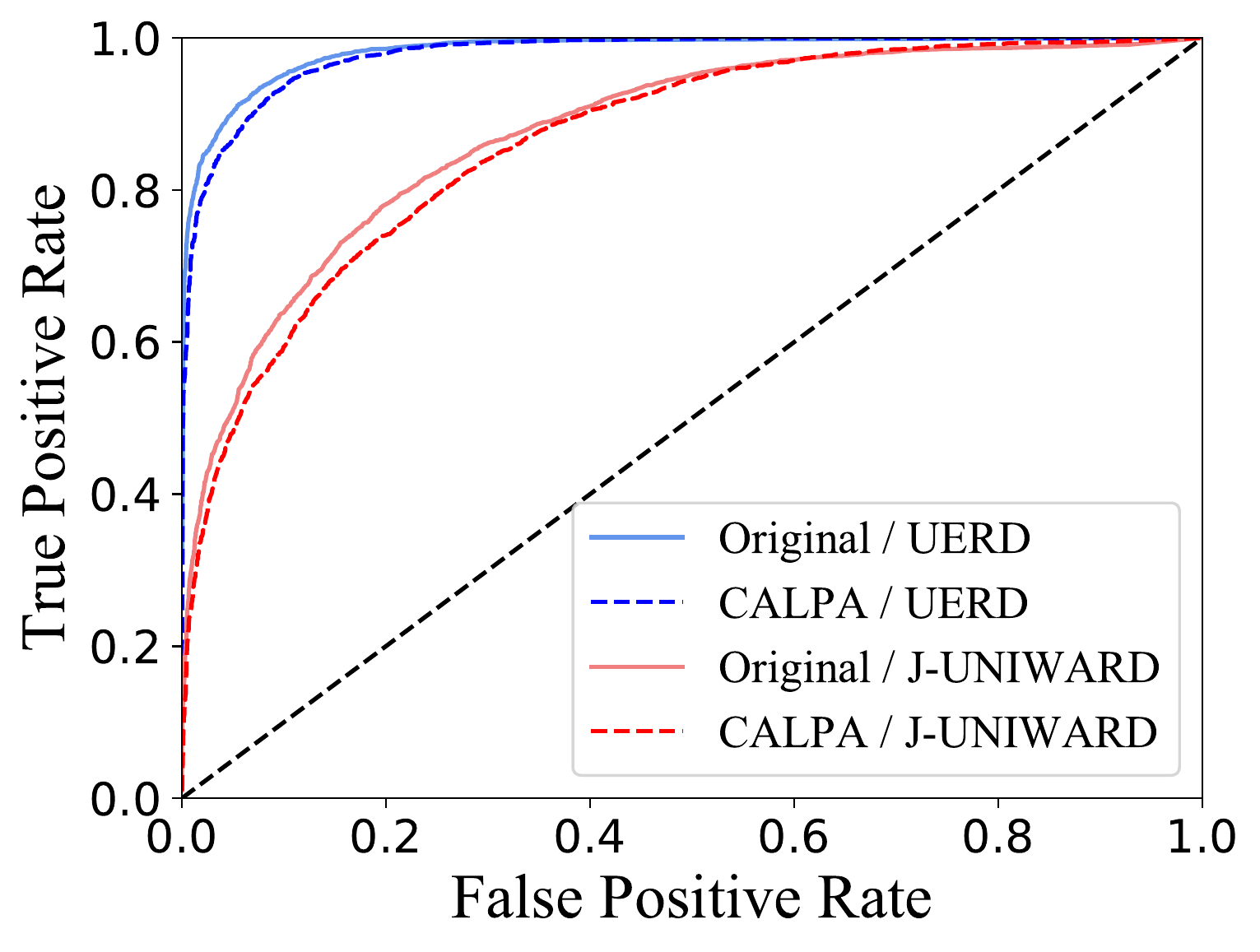} 
  }\hspace{.05\linewidth}
  \subfloat[]{
    \label{fig:srnet_spatial_hill_0204}
    \includegraphics[width=0.4\textwidth,keepaspectratio]{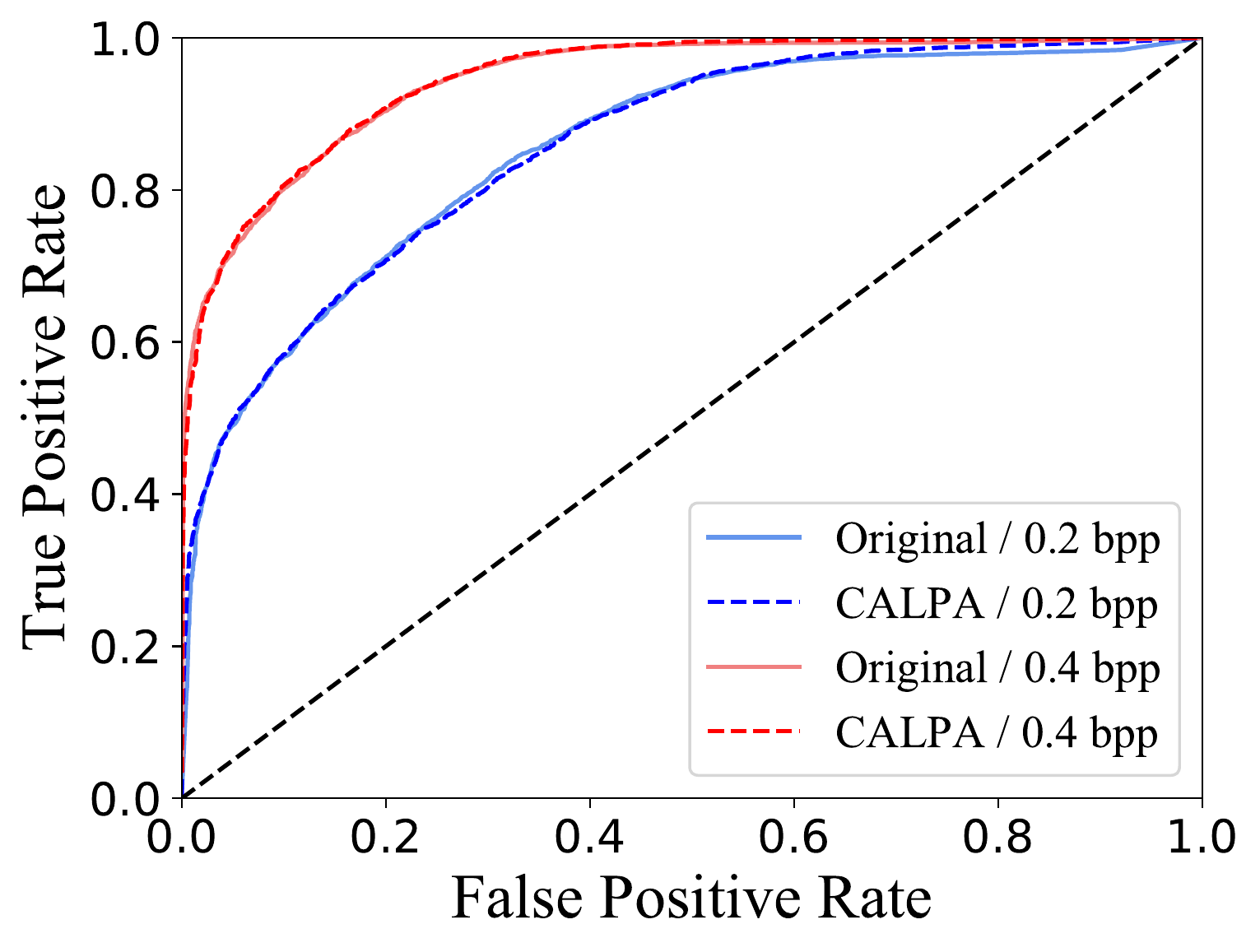}
  }\\
  \caption[]{Comparison of testing accuracy of CALPA-SRNet and the
    corresponding original
    SRNet. \subref{fig:srnet_jpg75_juniward_0204} The trained models
    are aiming at detecting J-UNIWARD stego images with QF 75.
    \subref{fig:srnet_jpg75_uerd_0204} Aiming at detecting UERD stego
    images with QF 75. \subref{fig:srnet_jpg95_uerd_juniward_04} Aiming
    at detecting UERD/J-UNIWARD stego images with 0.4 bpnzAC payload
    and QF 95.  \subref{fig:srnet_spatial_hill_0204} Aiming at
    detecting HILL spatial-domain stego images with 0.2/0.4 bpp
    payload.}
  \label{fig:comp_state_of_the_art_srnet}
\end{figure*}

In Fig.~\ref{fig:comp_fine_tuned_from_scratch}, we show how the
training accuracy and validation accuracy change with successive
training iterations for the original SRNet, the corresponding
CALPA-SRNet trained from scratch, and the pruned SRNet with
``training-pruning-finetuning'' pipeline. When trained to detect
J-UNIWARD stego images with 0.4 bpnzAC payload and QF 75, SRNet
achieved the best validation accuracy at $42.35 \times 10^4$
iterations. The corresponding model was selected out to apply
channel-pruning-assisted search to get CALPA-SRNet.  CALPA-SRNet was
reset and trained from scratch. As shown in
Fig.~\ref{fig:comp_fine_tuned_from_scratch}, the validation accuracies
of CALPA-SRNet trained from scratch had become stable at around
$40 \times 10^4$ iterations. For comparison, training procedure of the
pruned SRNet was resumed. It was further finetuned till
$80 \times 10^4$ iterations. But it can be observed that the
validation accuracies of pruned SRNet during the finetuning procedure
was unstable. Furthermore, its best validation accuracy did not
surpass CALPA-SRNet trained from scratch.

\subsection{Detection performance of CALPA-NET}
\label{sec:comp_to_sta}

\begin{figure*}[!t]
  \centering
  % https://tex.stackexchange.com/questions/354307/centering-wide-2x2-figure-using-adjustbox
  \begin{adjustbox}{center}
    \subfloat[]{
      \label{fig:srnet_thinet_rates}
      \includegraphics[height=0.19\textheight,keepaspectratio]{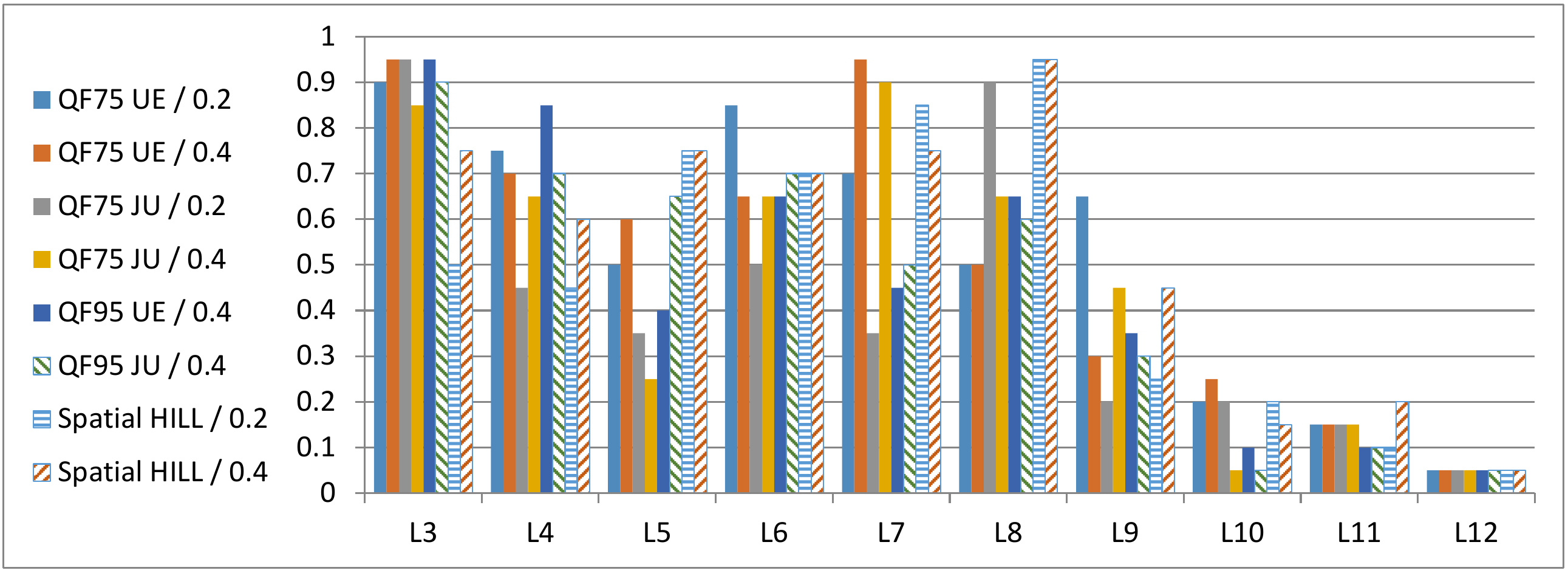} 
    }
    \subfloat[]{
      \label{fig:srnet_l1_rates}
      \includegraphics[height=0.19\textheight,keepaspectratio]{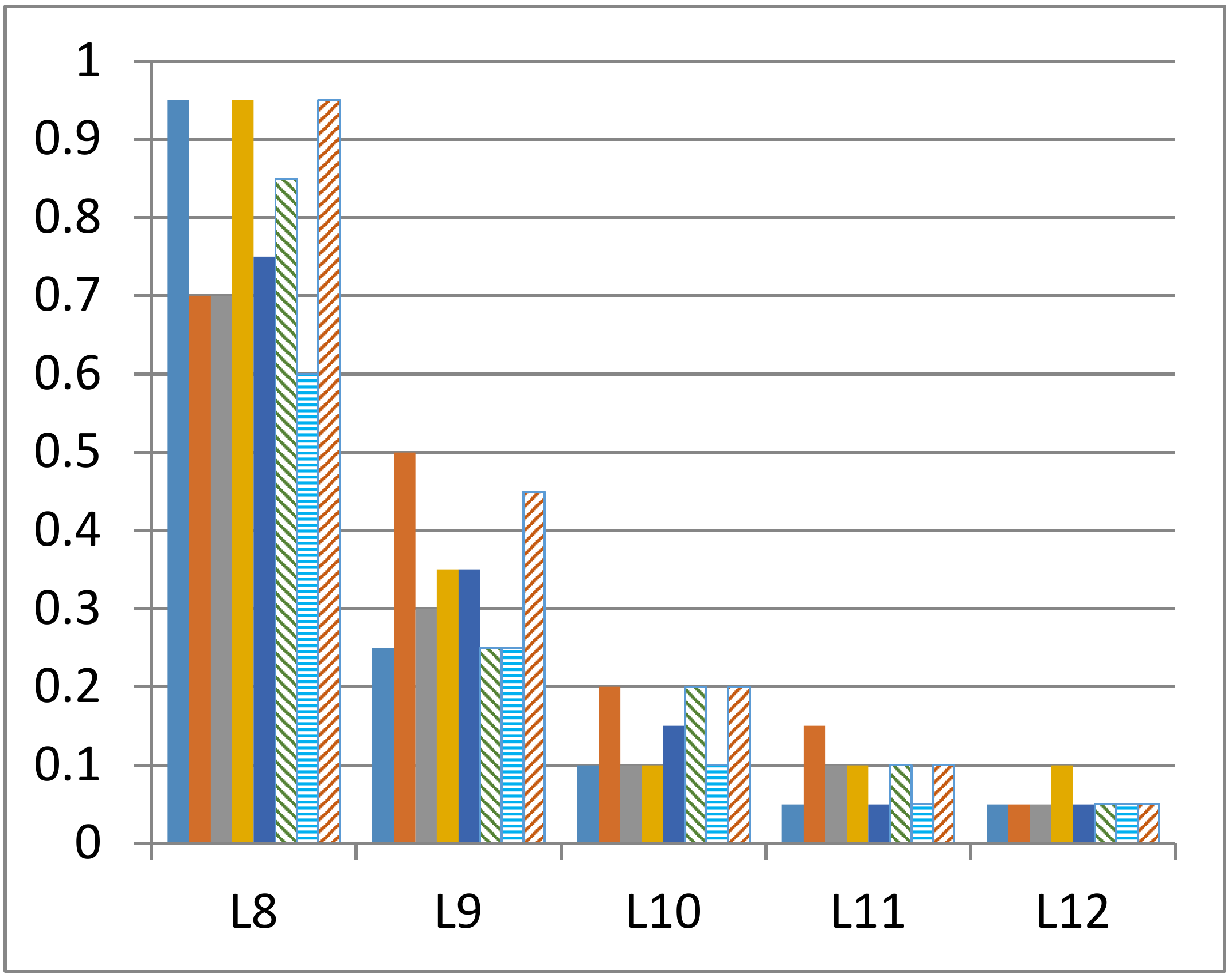} 
    }
  \end{adjustbox}
  \caption[]{The corresponding shrinking rates of CALPA-SRNets used in
    Fig.~\ref{fig:comp_state_of_the_art_srnet}. \subref{fig:srnet_thinet_rates}
    is for ThiNet scheme in blocks
    ``L3''---``L12''. \subref{fig:srnet_l1_rates} is for $l_1$-norm
    based scheme in transformed shortcut connections of blocks
    ``L8''---``L12''.}
  \label{fig:srnet_thinet_l1_rates}
\end{figure*}

\begin{table*}[!t]
  \centering
  \caption[]{Comparison of detection performance of CALPA-XuNet2 and
    the corresponding original XuNet2. The detection performance is
    measured via false-alarm rate for a given stego-image detection
    probability. Results for J-UNIWARD/UERD stego images with 0.2/0.4
    bpnzAC payload and QF 75/95 are included.}
  \resizebox{0.9\textwidth}{!}{%
    {\renewcommand{\arraystretch}{1.5}% for the vertical padding
      \begin{tabular}{c|cc|ccc|ccc}
        \Xhline{2\arrayrulewidth}
        & \multicolumn{2}{ c|}{} & \multicolumn{3}{ c| }{\textbf{CALPA-XuNet2}} & \multicolumn{3}{ c }{\textbf{XuNet2}} \\
        \hline
        {\textbf{Quality Factors}} & \multicolumn{2}{ c|}{\textbf{Targets}} & $P_{FA}(70\%)$ & $P_{FA}(50\%)$ & $P_{FA}(30\%)$ & $P_{FA}(70\%)$ & $P_{FA}(50\%)$ & $P_{FA}(30\%)$ \\
        \hline
        \multirow{4}{*}{\textbf{QF75}} & \multirow{2}{*}{\textbf{J-UNIWARD}} & 0.2 bpnzAC & 1.54\% & 0.30\% & 0.12\% & 1.06\% & 0.13\% & 0\% \\
        & & 0.4 bpnzAC & 0.1\% & 0.02\% & 0.01\% & 0.15\% & 0.04\% & 0.02\% \\
        & \multirow{2}{*}{\textbf{UERD}} & 0.2 bpnzAC & 0.91\% & 0.12\% & 0.02\% & 0.7\% & 0.14\% & 0.04\% \\
        & & 0.4 bpnzAC & 0\% & 0\% & 0\% & 0.07\% & 0\% & 0\% \\
        \hline
        \multirow{4}{*}{\textbf{QF95}} & \multirow{2}{*}{\textbf{J-UNIWARD}} & 0.2 bpnzAC & 36.16\% & 17.64\% & 3.28\% & 41.36\% & 22.69\% & 7.06\% \\
        & & 0.4 bpnzAC & 1.06\% & 0.06\% & 0\% & 0.72\% & 0.06\% & 0\%\\
        & \multirow{2}{*}{\textbf{UERD}} & 0.2 bpnzAC & 6.62\% & 0.66\% & 0.02\% & 4.64\% & 0.54\% & 0.06\% \\
        & & 0.4 bpnzAC & 0.46\% & 0.09\% & 0.02\% & 0.46\% & 0.10\% & 0.05\% \\
        \Xhline{2\arrayrulewidth}
      \end{tabular} 
    }} 
  \label{tab:comp_state_of_the_art_xunet}
\end{table*}

In Fig.~\ref{fig:comp_state_of_the_art_srnet}, we compare the
detection performance of CALPA-SRNet and the corresponding original
SRNet. From
Fig.~\ref{fig:comp_state_of_the_art_srnet}\subref{fig:srnet_jpg75_juniward_0204}
and
Fig.~\ref{fig:comp_state_of_the_art_srnet}\subref{fig:srnet_jpg75_uerd_0204}
we can see that CALPA-SRNet obtains comparable detection performance
when aiming at detecting JPEG stego images with QF 75~(drops a bit
with J-UNIWARD/0.2 bpnzAC). The detection performance of CALPA-SRNet
is slightly worse than original SRNet when JPEG stego images are with
QF 95, as shown in
Fig.~\ref{fig:comp_state_of_the_art_srnet}\subref{fig:srnet_jpg95_uerd_juniward_04}. Fig.~\ref{fig:comp_state_of_the_art_srnet}\subref{fig:srnet_spatial_hill_0204}
indicates that when used to detect 0.4 bpp HILL spatial-domain stego
images, CALPA-SRNet can also obtain almost the same detection
performance. Please note that all the above high performance of
CALPA-SRNet are achieved with mere 1\%$\sim$3\% parameters compared to
original SRNet.

In Fig.~\ref{fig:srnet_thinet_l1_rates}, we show the corresponding
shrinking rates of CALPA-SRNets used in
Fig.~\ref{fig:comp_state_of_the_art_srnet}. From
Fig.~\ref{fig:srnet_thinet_l1_rates} we can observe that the shrinking
rate of every convolutional layer adapts to the target steganographic
scheme, the embedding payload, the embedding domain and even the
actual quality factor used in JPEG target images. In general, the
shrinking rate reduces as the level of the corresponding convolutional
layer gets higher and higher. Significantly low shrinking rates can
always be observed in top convolutional layers.

In Tab.~\ref{tab:comp_state_of_the_art_xunet}, we compare the
detection performance of CALPA-XuNet2 and original XuNet2. We adopt an
alternative performance measure used in \cite{boroumand_tifs_2018}:
the false-alarm rate for a given stego-image detection
probability. For instance, $P_{FA}(30\%)$ denotes the false-alarm rate
for 30\% stego-image detection probability. From
Tab.~\ref{tab:comp_state_of_the_art_xunet}, by looking into all the
false-alarm rates with stego-image detection probability stepping from
30\% to 50\% and then to 70\%, it is clear that on the whole, there is
no distinction between the detection performance of CALPA-XuNet2 and
that of XuNet2. The results demonstrate the amazing detection
performance of CALPA-XuNet2 since it is quite a lightweight model
compared to original XuNet2. The architecture of one of the
CALPA-XuNet2 models aiming at detecting J-UNIWARD stego images with
0.4 bpnzAC payload and QF 75 can be found in
Fig.~\ref{fig:calpa_xunet_srnet_structure}, in which we can see
CALPA-XuNet2 also presents a slender bottleneck-like structure.

\subsection{Adaptivity, Transferability, and Scalability of CALPA-NET}
\label{sec:transferability}

\begin{figure*}[!t]
  \centering
  \subfloat[]{
    \label{fig:comp_calpa_avg_aggr_juniward}
    \includegraphics[width=0.4\textwidth,keepaspectratio]{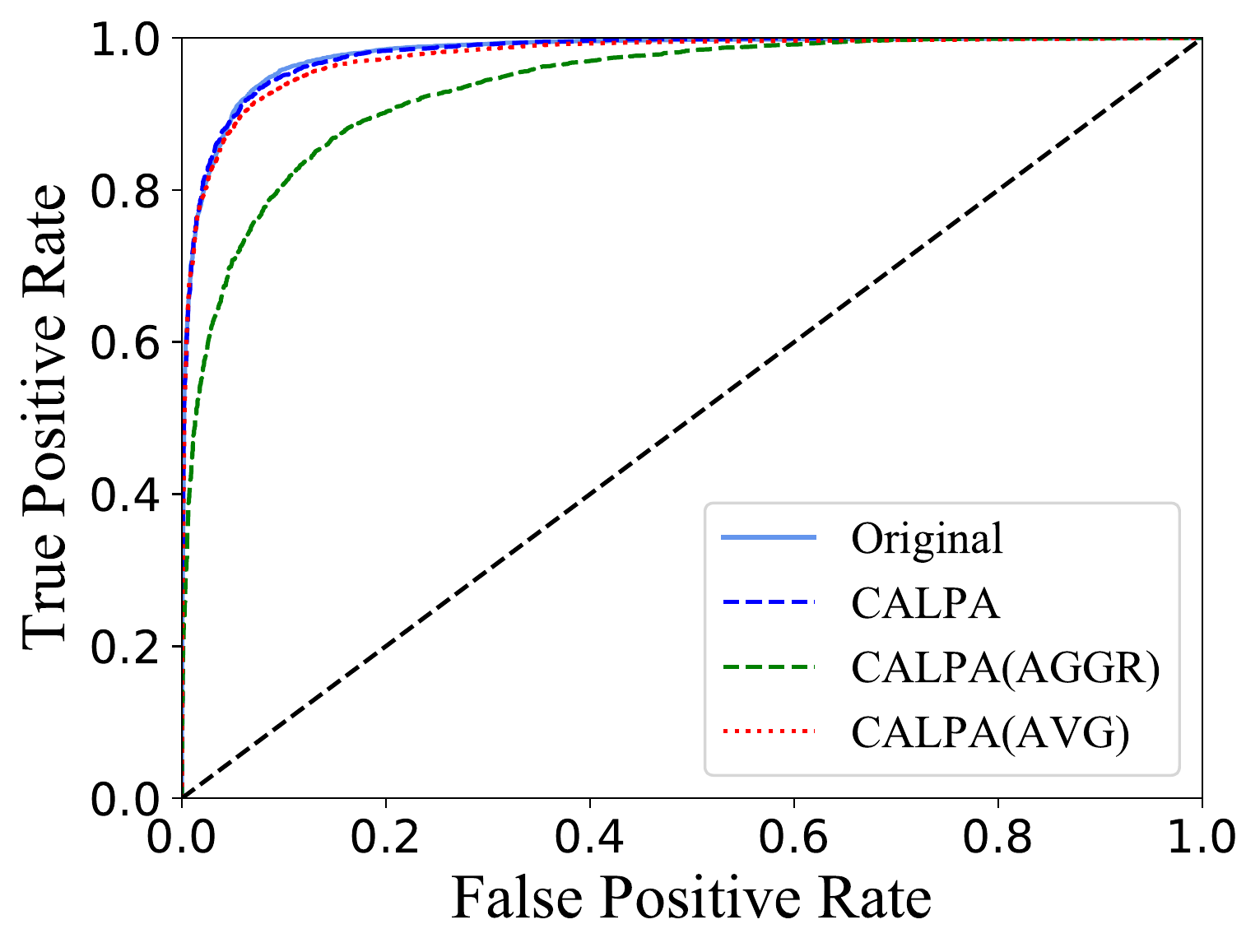} 
  }\hspace{.05\linewidth} 
  \subfloat[]{
    \label{fig:comp_calpa_avg_aggr_hill}
    \includegraphics[width=0.4\textwidth,keepaspectratio]{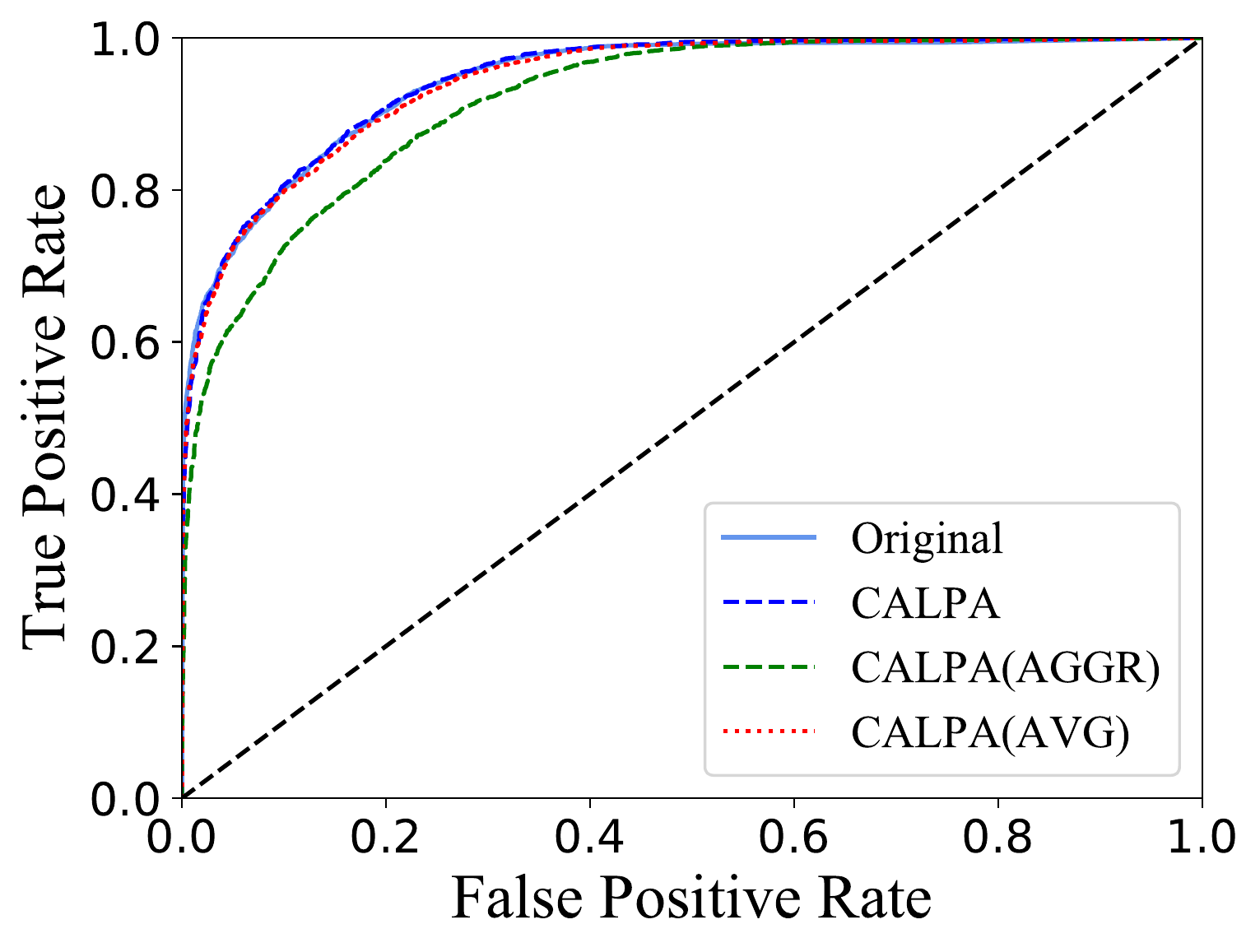}
  }\\
  \caption[]{Comparison of testing accuracy of CALPA-SRNet with two less adaptive
    alternative schemes, the AGGR scheme and the AVG
    scheme. \subref{fig:comp_calpa_avg_aggr_juniward} For J-UNIWARD
    stego images with 0.4 bpnzAC payload and QF
    75. \subref{fig:comp_calpa_avg_aggr_hill}For HILL stego images with
    0.4 bpp payload.} 
  \label{fig:comp_calpa_avg_aggr}
\end{figure*} 

\begin{figure*}[!t]
  \centering
  \subfloat[]{
    \label{fig:transfer_pruned_structure_juniward}
    \includegraphics[width=0.4\textwidth,keepaspectratio]{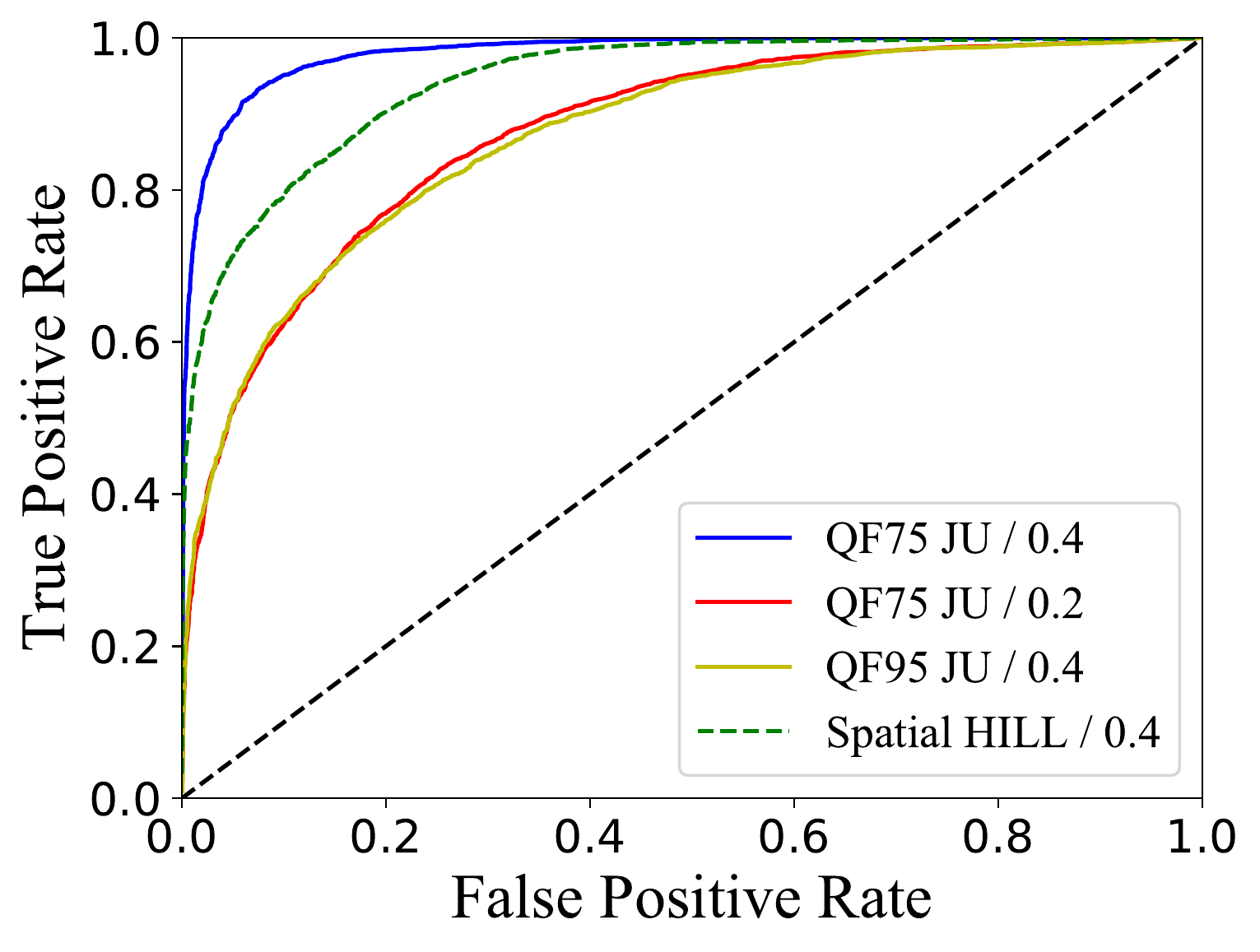} 
  }\hspace{.05\linewidth}  
  \subfloat[]{
    \label{fig:transfer_pruned_structure_hill}
    \includegraphics[width=0.4\textwidth,keepaspectratio]{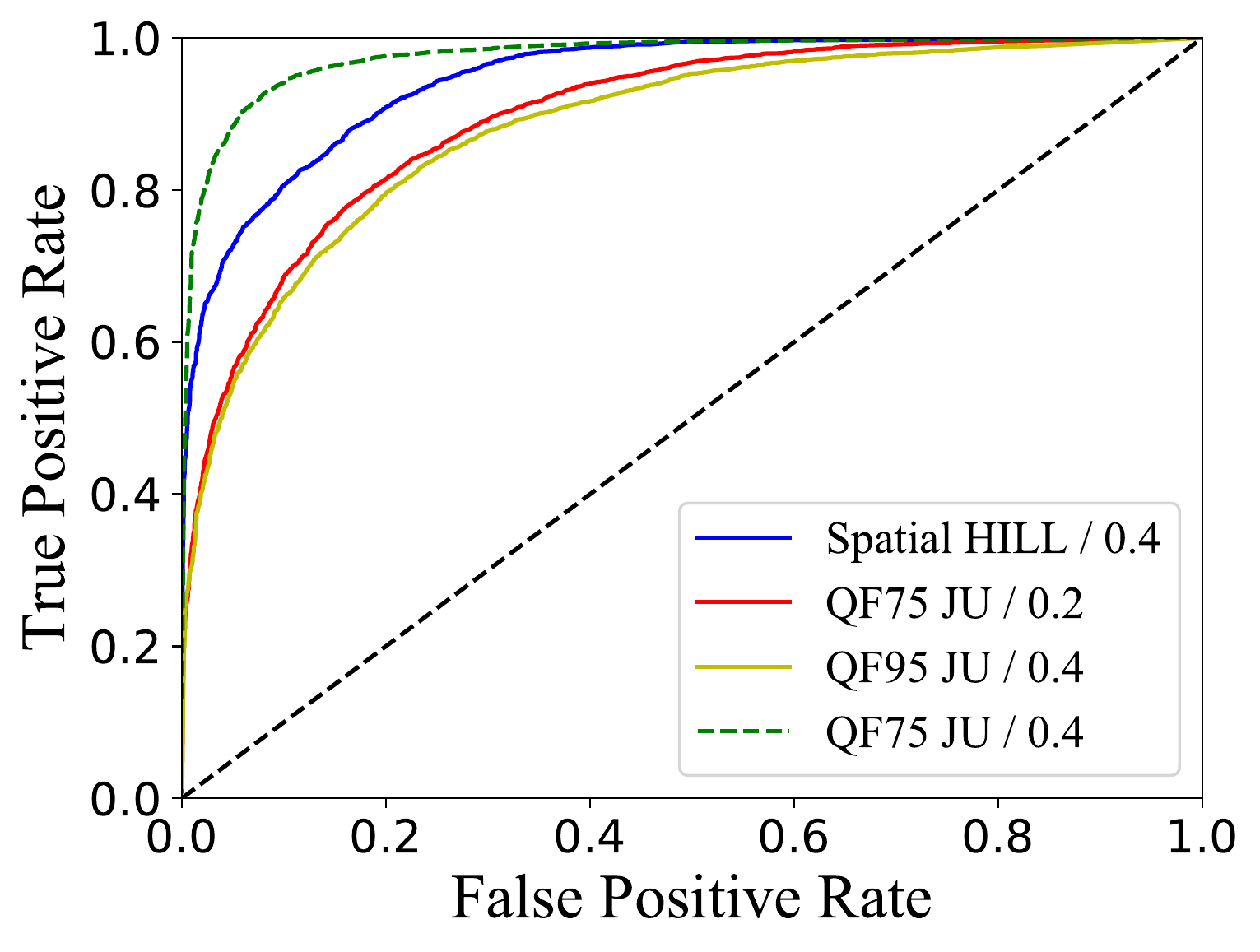}   
  }\\
  \caption[]{Detection performance of CALPA-SRNet when applied to
    another targets. ``JU'' in the legend is short for J-UNIWARD
    steganographic
    algorithm. \subref{fig:transfer_pruned_structure_juniward} The
    architecture is original aiming at J-UNIWARD stego images with 0.4
    bpnzAC payload and QF
    75. \subref{fig:transfer_pruned_structure_hill} The architecture is
    original aiming at HILL stego images with 0.4 bpp payload. }
  \label{fig:transfer_pruned_structure}
\end{figure*} 

As demonstrated in prior experiments, the structure of CALPA-NET is
adaptive to the actual targets it aims at due to the fact that
CALPA-NET is data-driven. Firstly in
Fig.~\ref{fig:comp_calpa_avg_aggr}, we evaluate the effectiveness of
the adaptive data-driven scheme used in CALPA-NET taking CALPA-SRNet
as example. Here two less adaptive alternative schemes are introduced:
\begin{itemize}
\item AGGR scheme: The global shrinking rate of all the inclusive
  convolutional layers is aggressively set to the minimal one
  determined in the CALPA-NET bottom-up traversal.
\item AVG scheme: The global shrinking rate of all the inclusive
  convolutional layers is set to the average of all the determined
  shrinking rates.
\end{itemize}
Refer back to the CALPA-SRNet illustrated in
Fig.~\ref{fig:calpa_xunet_srnet_structure}, the global shrinking rate
can be obtained as 5\% and 56\% for AGGR scheme and AVG scheme
respectively. From Fig.~\ref{fig:comp_calpa_avg_aggr} we can see
though AGGR scheme has the least parameters and FLOPs, its detection
performance degrades remarkably. AVG scheme possesses similar scale of
parameters and FLOPs as CALPA-SRNet, but no longer has bottleneck-like
structure. Its detection performance is close to CALPA-SRNet. However,
it is still a bit inferior to CALPA-SRNet when aiming at J-UNIWARD
stego images.

Next, we investigate the transferability of the obtained architecture
of CALPA-NET.  Firstly we take the CALPA-SRNet architecture aiming at
J-UNIWARD stego images with 0.4 bpnzAC payload and QF 75 as
target. From
Fig.~\ref{fig:transfer_pruned_structure}\subref{fig:transfer_pruned_structure_juniward}
it is clear that the CALPA-SRNet architecture obtains good detection
performance in the three mismatched scenarios.  Furthermore, the
specific JPEG-oriented CALPA-SRNet architecture remains effective for
spatial-domain HILL stego images, implying that HILL and J-UNIWARD
share some intrinsic characteristics although they are in completely
different domain. Oppositely, we also take the CALPA-SRNet
architecture aiming at HILL stego images in spatial domain with 0.4
bpp payload as
target. Fig.~\ref{fig:transfer_pruned_structure}\subref{fig:transfer_pruned_structure_hill}
shows its detection performance on the three JPEG domain scenarios. We
can see that the CALPA-SRNet architecture originally for spatial
domain scenario is still effective on the three diverse JPEG domain
scenarios, which implies that our proposed CALPA-NET approach presents
a cross-domain universal characteristics.

\begin{table*}[!t]
  \centering
  \caption[]{Detection performance of CALPA-SRNet and original SRNet
    trained on one target and then tested on another target. The
    payload of the targets were fixed to 0.4 bpnzAC/bpp. The detection
    performance was measured with total error probability $P_E$.}
  \label{tab:transfer_trained_model}
  {\renewcommand{\arraystretch}{1.5}% for the vertical padding
   \begin{tabular}[c]{|l|c|@{\hspace{3em}}c@{\hspace{3em}}|c|c|@{\hspace{3em}}c@{\hspace{3em}}|}
     \hline
     \multicolumn{3}{|c|}{\textbf{JPEG domain}} & 
     \multicolumn{3}{c|}{\textbf{Spatial domain}}  \\
     \hline
     \hline
     \multicolumn{6}{|c|}{\textbf{CALPA-SRNet}}\\
     \hline
     \diagbox[width=9.5em]{Trained on}{Tested on}
       & J-UNIWARD & UERD &
     \diagbox[width=9.5em]{Trained on}{Tested on}
       & S-UNIWARD & HILL \\
     \hline
     J-UNIWARD & 7.5\% & 5.64\% & S-UNIWARD & 10.63\% & 26.03\% \\
     \hline
     UERD & 24.2\% & 3.17\% &HILL & 23.89\% & 14.63\% \\
     \hline
     \multicolumn{6}{|c|}{\textbf{SRNet}}\\
     \hline
     \diagbox[width=9.5em]{Trained on}{Tested on}
       & J-UNIWARD & UERD &
     \diagbox[width=9.5em]{Trained on}{Tested on}
       & S-UNIWARD & HILL \\
     \hline
     J-UNIWARD & 7.02\% & 5.35\% & S-UNIWARD & 12.75\% & 26.99\% \\
     \hline
     UERD & 26.04\% & 3.37\% & HILL & 22.48\% & 14.7\% \\
     \hline
   \end{tabular}
  }
\end{table*}

We then investigate the transferability of the trained CALPA-NET
model. In Tab.~\ref{tab:transfer_trained_model} we compare the
detection performance of CALPA-SRNet and original SRNet trained on one
target and tested on another target. In order to make a fair
comparison, we used the following performance measurement~(as in
Tab.~II of \cite{boroumand_tifs_2018}):
$P_E=\min_{P_{FA}}\frac{1}{2}(P_{FA}+P_{MD})$, where $P_{FA}$ and
$P_{MD}$ are the false alarm rate and miss detection rate. From
Tab.~\ref{tab:transfer_trained_model} it can be clearly observed that
when the payload of the targets are the same, CALPA-SRNet achieves
similar, and even better transferability compared with original SRNet.

\begin{figure*}[!t]
  \centering
  \subfloat[]{
    \label{fig:calpa_srnet_large_scale_juniward}
    \includegraphics[width=0.4\textwidth,keepaspectratio]{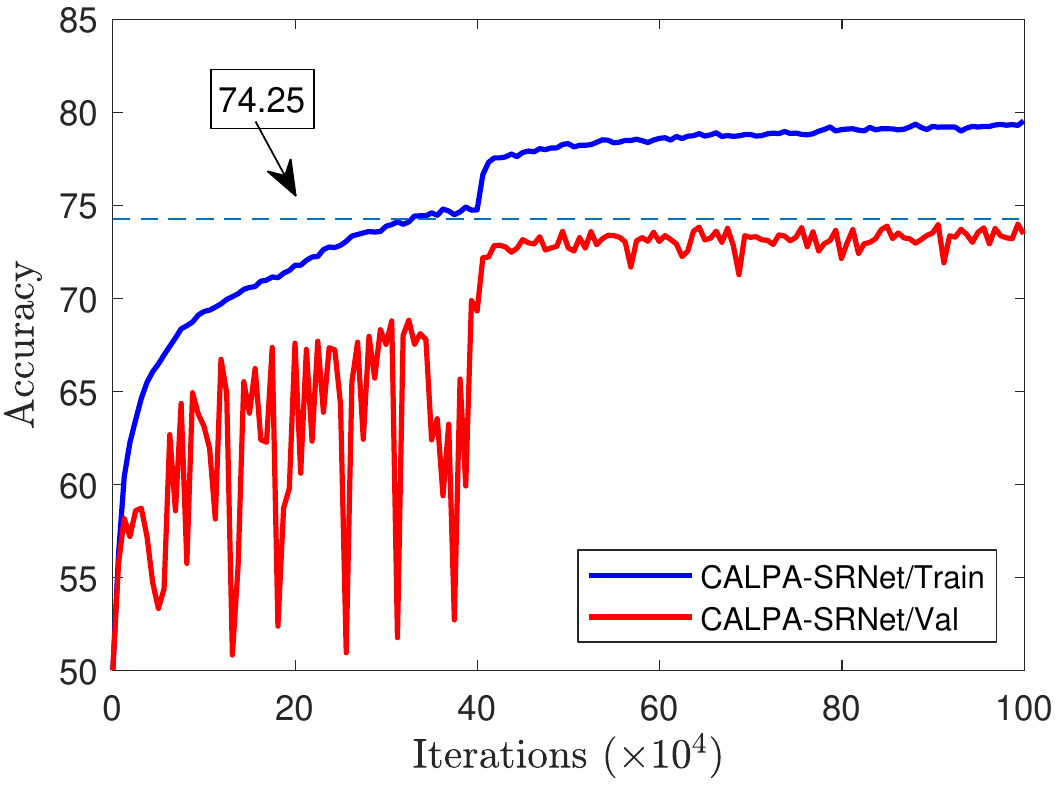}
  }\hspace{.05\linewidth}
  \subfloat[]{
    \label{fig:srnet_large_scale_juniward}
    \includegraphics[width=0.4\textwidth,keepaspectratio]{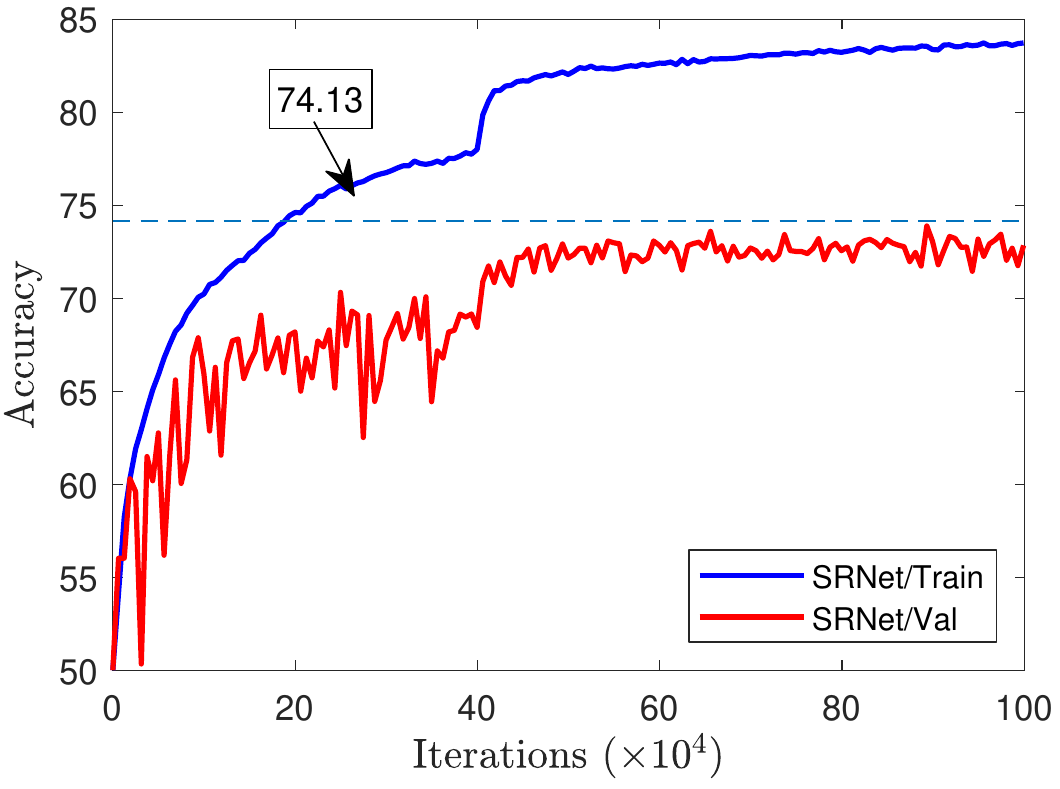} 
  }\\
  \subfloat[]{
    \label{fig:calpa_srnet_large_scale_uerd}
    \includegraphics[width=0.4\textwidth,keepaspectratio]{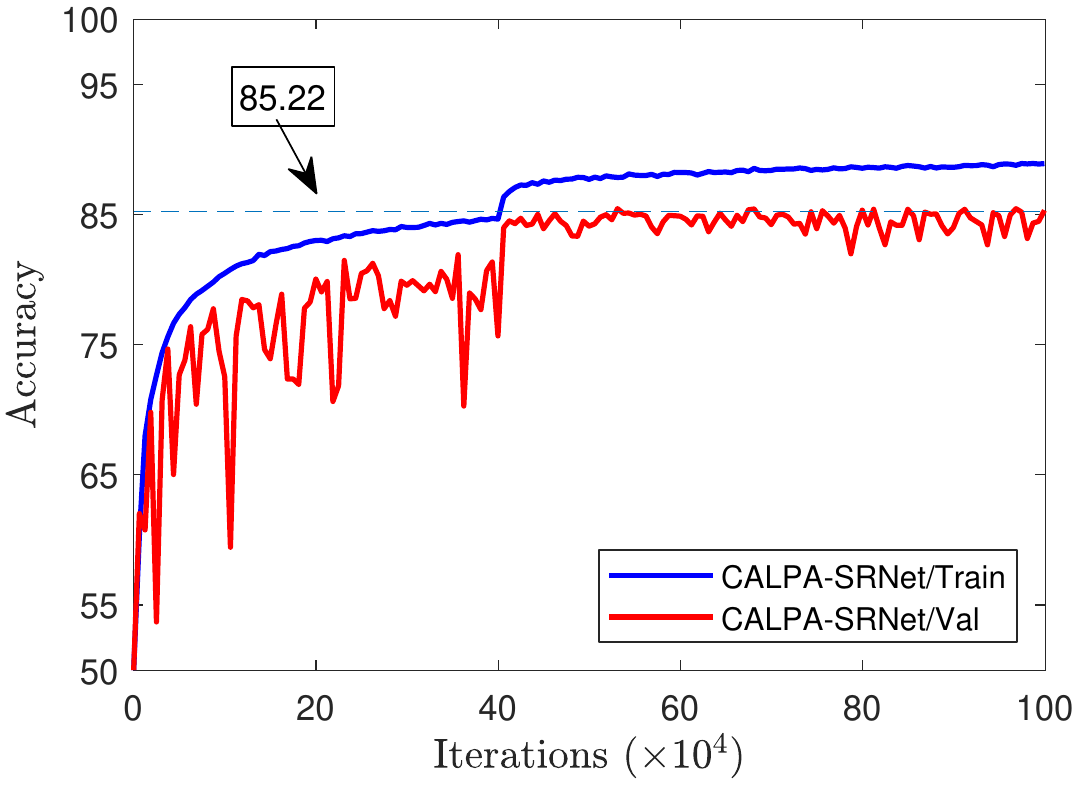}
  }\hspace{.05\linewidth}
  \subfloat[]{
    \label{fig:srnet_large_scale_uerd}
    \includegraphics[width=0.4\textwidth,keepaspectratio]{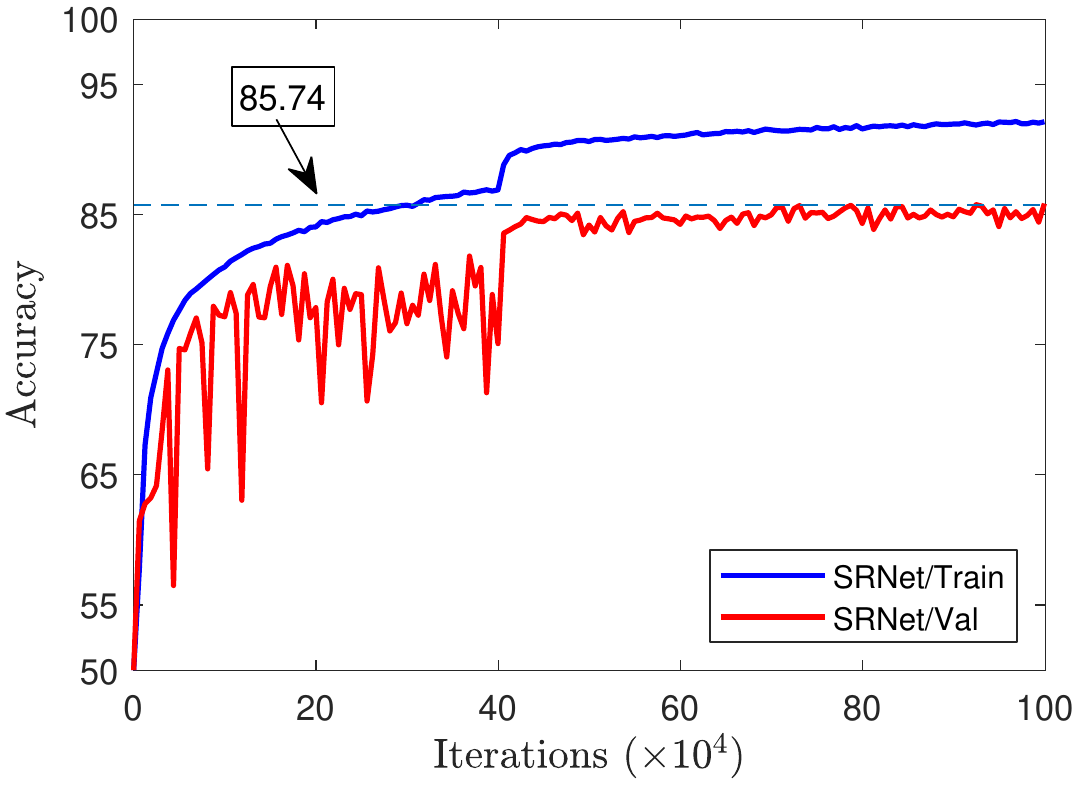} 
  }\\ 
  \caption[]{ Comparison of the training accuracies, validation
    accuracies and testing accuracies vs. training iterations for
    CALPA-SRNet and the corresponding original SRNet on a subset of
    CLS-LOC dataset. The trained models are aiming at detecting
    J-UNIWARD/UERD stego images with 0.4 bpnzAC payload and QF 75. The
    blue dashed line in every sub-figure marks the highest testing
    accuracy achieved by the trained
    model. \subref{fig:calpa_srnet_large_scale_juniward} CALPA-SRNet,
    aiming at J-UNIWARD; \subref{fig:srnet_large_scale_juniward}
    Original SRNet, aiming at J-UNIWARD;
    \subref{fig:calpa_srnet_large_scale_uerd} CALPA-SRNet, aiming at
    UERD; \subref{fig:srnet_large_scale_uerd} Original SRNet, aiming at
    UERD.}
  \label{fig:comp_state_of_the_art_srnet_large_scale}
\end{figure*}

At last the scalability of CALPA-NET is investigated.  We firstly
demonstrate the performance of CALPA-SRNet on the subset of ImageNet
CLS-LOC dataset, which is of great cover source diversity. From
Fig.~\ref{fig:comp_state_of_the_art_srnet_large_scale} we can see even
in a tenfold larger dataset, CALPA-SRNet still shows similar
validation and testing performance compared to original SRNet. Please
note that the gap between training accuracies and validation
accuracies for original SRNet is much bigger than that for
CALPA-SRNet, indicating that CALPA-SRNet may be less vulnerable to
overfitting the training set.

\begin{figure*}[!t]
  \centering
  \subfloat[]{
    \label{fig:comp_fine_tuned_from_scratch_alaska}
    \raisebox{-0.3\height}{\includegraphics[width=0.82\textwidth,keepaspectratio]{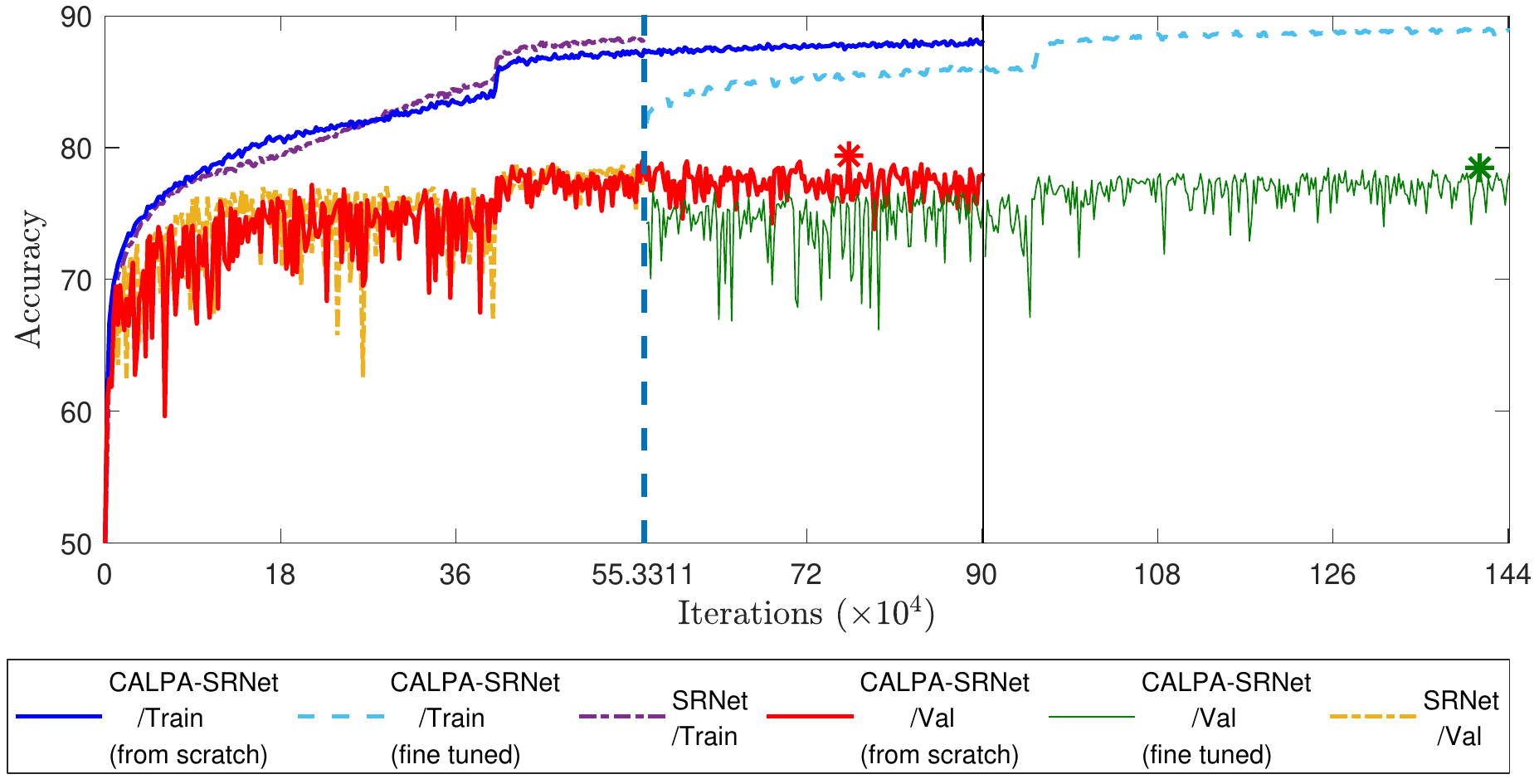}}
  }
  \subfloat[]{
    \label{fig:calpa_srnet_structure_alaska}
    \raisebox{-0.3\height}{\includegraphics[width=0.155\textwidth,keepaspectratio]{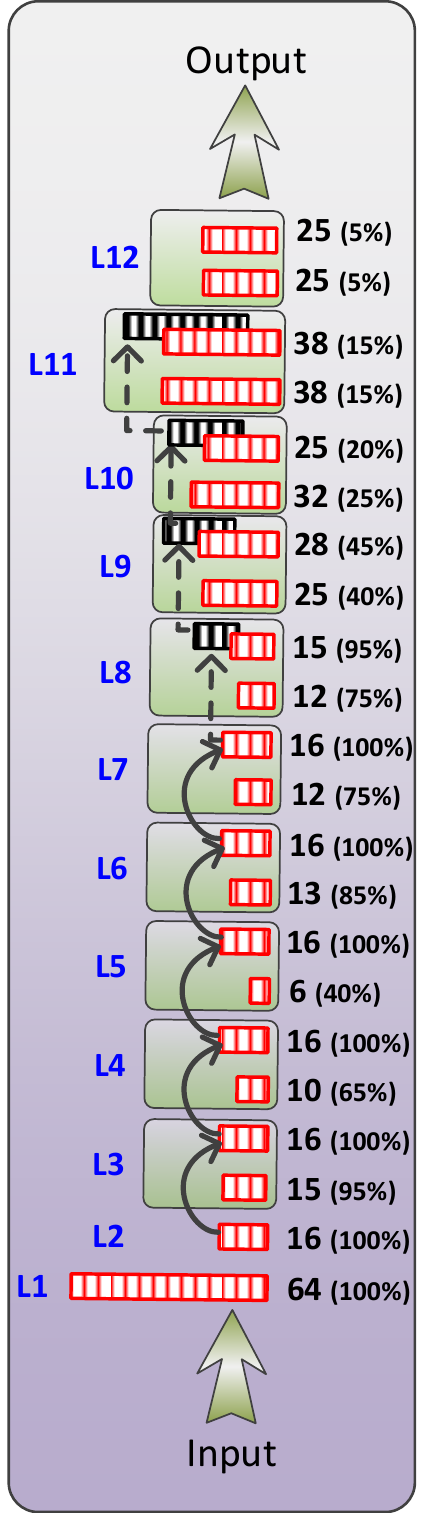}}
  }\\
  \caption[]{\subref{fig:comp_fine_tuned_from_scratch_alaska}
    Comparison of the training accuracies and validation accuracies
    vs. training iterations for the original SRNet, CALPA-SRNet
    trained from scratch, and CALPA-SRNet with
    ``training-pruning-finetuning'' pipeline on ALASKA
    dataset. ``\ding{107}'' denotes the point with the highest value
    at the corresponding
    plot. \subref{fig:calpa_srnet_structure_alaska} The conceptual
    structure of the corresponding CALPA-SRNet model.}
  \label{fig:alaska_results_and_diagram}  
\end{figure*} 

Then we evaluate the performance of CALPA-SRNet on ALASKA, which is of
great stego source diversity. For the 80,005 ALASKA images, the
train/validation/test dataset partition is
35,000/5,000/40,005. Following the configuration in
\cite{cogranne_ihmmsec_2019}, three more JPEG steganographic
algorithms, UED~\cite{guo_wifs_2012}, EBS~\cite{wang_icassp_2012} and
nsF5~\cite{fridrich_mmsec_2007} were used, alongside with
J-UNIWARD. The probability of using each of the steganographic
algorithms above was 30\%, 15\%, 15\% and 40\%, respectively. The
corresponding embedding rate was 0.25 bpnzAC, 0.25 bpnzAC, 0.05 bpnzAC
and 0.4 bpnzAC, respectively~\footnote{The default embedding rates in
  the ALASKA embedding script were adopted since the logs of
  processing pipelines have not been provided for ALASKA v2.}. 

From
Fig.~\ref{fig:alaska_results_and_diagram}\subref{fig:comp_fine_tuned_from_scratch_alaska},
we can see SRNet achieved the best validation accuracy at
$55.3311 \times 10^4$ iterations. The CALPA-SRNet architecture
obtained from it could become stable at around $40 \times 10^4$
iteration when trained from scratch. The highest validation accuracy
of CALPA-SRNet was at the same level as that of the original
SRNet. Compared to Fig.~\ref{fig:comp_fine_tuned_from_scratch}, we can
see that on ALASKA dataset, the validation accuracies of CALPA-SRNet
trained from scratch are obviously higher than the pruned SRNet with
traditional ``training-pruning-finetuning'' pipeline. For the sake of
completeness, the testing accuracies of the three models with the best
validation accuracy are also reported here---CALPA-SRNet: 78.73\%; the
pruned SRNet with ``training-pruning-finetuning'' pipeline: 77.53\%;
original SRNet: 78.42\%. 

The conceptual structure of the corresponding CALPA-SRNet model is
illustrated in
Fig.~\ref{fig:alaska_results_and_diagram}\subref{fig:calpa_srnet_structure_alaska}. Compared
with the conceptual structure of CALPA-SRNet in
Fig.~\ref{fig:calpa_xunet_srnet_structure}, we can see that though
there are certain differences, the CALPA-SRNet for ALASKA also
presents a slender bottleneck-like structure. As a result, compared
with original SRNet, it is only with $1.97\%$
parameters~(9.22$\times 10^4$) and $36.5\%$ FLOPs~(2.17$\times 10^9$).

\subsection{Further discussion: why train CALPA-NETs from scratch}
\label{sec:further-discussion}

Putting all the experimental evidences together, the reasons for
training CALPA-NETs from scratch are provided as follows:

Firstly, in view of detection performance, training CALPA-NETs from
scratch is better than finetuning it. In
Tab.~\ref{tab:threshold_candidates}, it is clear that the detection
performance of CALPA-SRNet trained from scratch is slightly better
than finetuning it on BOSSBase+BOWS2 dataset, when tolerable accuracy
lost $\varsigma$ is low. On ALASKA dataset, a larger and more diverse
dataset, the detection performance of CALPA-SRNet trained from scratch
is obviously better than the pruned SRNet with
``training-pruning-finetuning'' pipeline.

Secondly, in view of training efficiency, training CALPA-NETs from
scratch is better than finetuning
it. Fig.~\ref{fig:comp_fine_tuned_from_scratch} shows that on
BOSSBase+BOWS2 dataset, the validation accuracies of pruned SRNet
during the finetuning procedure was unstable. We can see more distinct
efficiency difference on the ALASKA dataset. As shown in
Fig.~\ref{fig:alaska_results_and_diagram}\subref{fig:comp_fine_tuned_from_scratch_alaska},
even the best validation obtained by the pruned SRNet at further
$85.7304 \times 10^4$
iterations~($(85.7304+55.3311)\times 10^4=141.0615\times 10^4$) was
inferior.  As comparison, the validation accuracies of CALPA-SRNet
trained from scratch became stable at around $40 \times 10^4$
iterations and achieved the peak at mere $76.3263 \times 10^4$
iterations.

Thirdly and the most importantly, the effectiveness of CALPA-NETs
trained from scratch demonstrates what matters most is the
cost-effective architecture obtained by CALPA-NET approach, rather
than the preserved parameters in the pruned steganalytic
models. Furthermore, Fig.~\ref{fig:transfer_pruned_structure}
demonstrates that CALPA-NET architecture presents general
applicability.
Fig.~\ref{fig:alaska_results_and_diagram}\subref{fig:comp_fine_tuned_from_scratch_alaska}
also demonstrates that there exists an effective slender
bottleneck-like CALPA-SRNet architecture under the stego source
diversity scenario. Therefore, the users can absolutely apply one
versatile CALPA-NET architecture to multiple steganalytic scenarios
and train it from scratch adaptively with limited budget. Such an
ability is impossible for traditional three-stage
training-pruning-finetuning pipeline.
\section{Concluding remarks}
\label{sec:conclude}

In this paper we propose CALPA-NET, which is aiming at adaptively
search efficient network structure on top of existing
over-parameterized and vast deep-learning based steganalyzers. The
major contributions of this work are as follows:
\begin{itemize}
\item We have observed that the broad inverted-pyramid structure of
  existing deep-learning based steganalyzers cannot boost detection
  performance even with fifty fold parameter. A theoretical reflection
  is proposed to argue that such structure might contradict the
  well-established model diversity oriented philosophy.
\item We have proposed a channel-pruning-assisted deep residual
  network architecture search approach which uses a hybrid criterion
  combined with ThiNet and $l_1$-norm based network pruning scheme. In
  the proposed network architecture, the traditional
  training-pruning-finetuning network pruning pipeline is completed
  abandoned.
\item The extensive experiments conducted on de-facto benchmarking
  image datasets show that CALAP-NET can achieve comparative detection
  performance with just a few percent of the model size and a small
  proportion of computational cost.
\end{itemize}

Our future work will focus on two aspects: (1) development of a fast
adaptive structural adjustment algorithm to make deep-learning based
steganalyzers self adapt to targets without the introduction of
redundant parameters/components; (2) further exploration of the
feasibility of completely automatic deep-learning based steganalytic
framework generation.

\appendices

\section{Theoretical reflection}
\label{app:reflection}

In the CNN pipeline, given a convolutional layer $L_l$, let us start
from \eqref{eq:conv}. Let
$\boldsymbol{\mathcal{O}}_{jk::}=\boldsymbol{\widehat{\mathcal{Z}}}^{l-1}_{j::}
\ast \boldsymbol{\mathcal{W}}^{l}_{jk::}$, then
$\boldsymbol{\mathcal{Z}}^{l}_{k::}=
\sum^{J}_{j=1}\boldsymbol{\mathcal{O}}_{jk::},\ 1 \le k \le K^{l}$.
Assume that $\boldsymbol{\mathcal{O}}_{jk::}$ is a discrete sample
drawn from a continuous latent distribution $\overline{O}_{jk}$, with
PDF~(Probability Distribution Function) $o_{jk}$. Accordingly, let
$\boldsymbol{\mathcal{Z}}^{l}_{k::}$ be a sample of the latent
distribution $\overline{Z}_k$ with PDF $z_k$. Please note that the PDF
of the aggregation is equal to the convolution of the PDF of the
aggregated terms:
\begin{equation}
  \label{eq:conv_in_spatial}
  z_k=\underbrace{o_{1k}\ast o_{2k} \ast \cdots \ast o_{jk} \ast \cdots \ast o_{Jk}}_{1 \le j \le J}
\end{equation}
Denote the Fourier transform of $o_{jk}$ and $z_k$ as
$\mathpzc{O}_{jk}$ and $\mathpzc{Z}_k$, respectively. Therefore in
frequency domain, we can get:
\begin{equation}
  \label{eq:conv_in_frequency}
  \mathpzc{Z}_{\,k}(\omega)=\prod_{j=1}^{J}\mathpzc{O}_{jk}(\omega)
\end{equation}

Given two frequencies $\omega_1$ and $\omega_2$, assume
that for a fixed $k$, $\mathpzc{O}_{jk}$ in $\omega_1$ contains sparse
intensity fluctuations, while in $\omega_2$ contains strong
intensities. Therefore for most $j$,
$|\mathpzc{O}_{jk}(\omega_1)| \ll
|\mathpzc{O}_{jk}(\omega_2)|$. Obviously, for large enough $J$:
\begin{equation}
  \label{eq:remove_fluctuations}
  \frac{|\mathpzc{Z}_{\,k}(\omega_1)|}{|\mathpzc{Z}_{\,k}(\omega_2)|}=\frac{\left|
    \prod_{j=1}^{J}\mathpzc{O}_{jk}(\omega_1)\right|}{\left|
    \prod_{j=1}^{J}\mathpzc{O}_{jk}(\omega_2)\right|}=
  \frac{\prod_{j=1}^{J}|\mathpzc{O}_{jk}(\omega_1)|}{\prod_{j=1}^{J}|\mathpzc{O}_{jk}(\omega_2)|}
  \longrightarrow 0
\end{equation}
From \eqref{eq:remove_fluctuations} we can see the aggregation in
\eqref{eq:conv} with large enough $J$ suppresses sparse intensity
fluctuations as well as heightens strong intensities in low-frequency
subbands of $\boldsymbol{\mathcal{Z}}^{l}_{k::}$, $1 \le k \le K^{l}$. 

% IEEEtran.bst should be put in the same directory as the source
% file. --tansq
\bibliographystyle{IEEEtran}
\bibliography{IEEEfull,psm_final}

\begin{IEEEbiography}[{\includegraphics[width=1in,height=1.25in,clip,keepaspectratio]{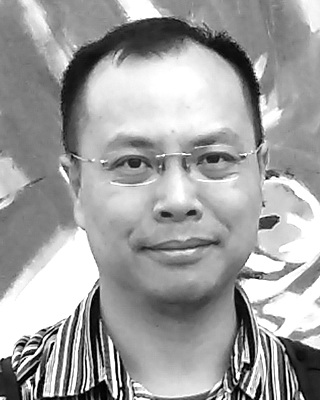}}]{Shunquan Tan (M'10--SM'17)}
  received the B.S. degree in computational mathematics and applied
  software and the Ph.D. degree in computer software and theory from
  Sun Yat-sen University, Guangzhou, China, in 2002 and 2007,
  respectively.

  He was a Visiting Scholar with New Jersey Institute of Technology,
  Newark, NJ, USA, from 2005 to 2006. He is currently an Associate
  Professor with College of Computer Science and Software Engineering,
  Shenzhen University, China, which he joined in 2007. His current
  research interests include multimedia security, multimedia
  forensics, and machine learning.
\end{IEEEbiography} 

\begin{IEEEbiography}[{\includegraphics[width=1in,height=1.25in,clip,keepaspectratio]{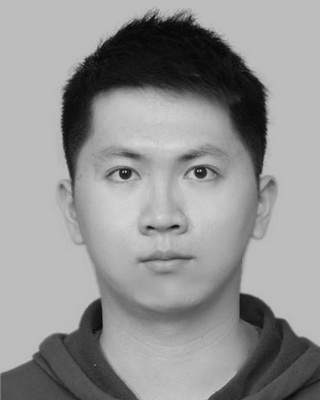}}]{Weilong
    Wu} received the B.S. degree of computer science and technology
  from Guangdong Ocean University, Zhanjiang, China in 2018. He is
  currently a master student in Shenzhen University majoring in
  software engineering. His current research interests include
  multimedia forensics, model compression, neural architecture search
  and deep learning.
\end{IEEEbiography}

\begin{IEEEbiography}[{\includegraphics[width=1in,height=1.25in,clip,keepaspectratio]{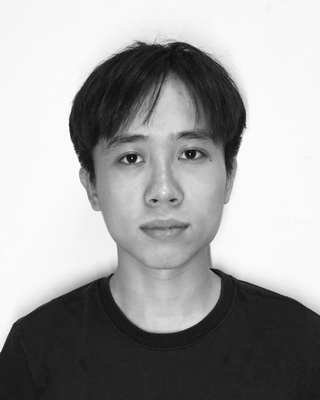}}]{Zilong Shao}
  received the B.S. degree of computer science and technology 
  from Shenzhen University, Shenzhen, China in 2019. He is 
  currently a master student in Shenzhen University majoring in
  computer technology. His current research interests include
  multimedia forensics, model compression, neural architecture 
  search and deep learning.
\end{IEEEbiography} 

\begin{IEEEbiography}[{\includegraphics[width=1in,height=1.25in,clip,keepaspectratio]{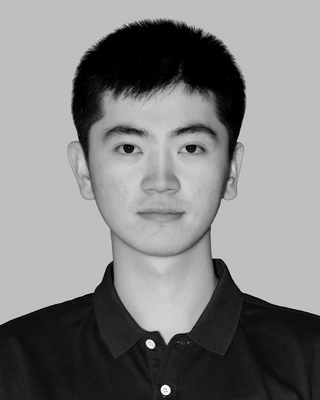}}]{Qiushi Li}
received the B.S. degree in information and computing science
from Harbin University of Science and Technology, Harbin, China,
in 2018. He is currently pursuing the M.S. degree in computer
science and technology with Shenzhen University, China. His
current research interests include multimedia forensics, model
compression and deep learning.
\end{IEEEbiography}

\begin{IEEEbiography}[{\includegraphics[width=1in,height=1.25in,clip,keepaspectratio]{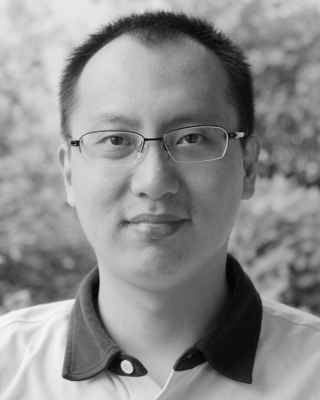}}]{Bin Li (S'07-M'09-SM'17)}
  received the B.E. degree in communication engineering and the
  Ph.D. degree in communication and information system from Sun
  Yat-sen University, Guangzhou, China, in 2004 and 2009,
  respectively.

  He was a Visiting Scholar with New Jersey Institute of Technology,
  Newark, NJ, USA, from 2007 to 2008. He is currently a Professor with
  Shenzhen University, Shenzhen, China, where he joined in 2009. He is
  also the director of Shenzhen Key Laboratory of Media Security, the
  vice director of Guangdong Key Lab of Intelligent Information
  Processing, and a member of Peng Cheng Laboratory. He has served as
  the IEEE Information Forensics and Security TC member since
  2019. His current research interests include multimedia forensics,
  image processing, and deep machine learning.
\end{IEEEbiography}

\begin{IEEEbiography}[{\includegraphics[width=1in,height=1.25in,clip,keepaspectratio]{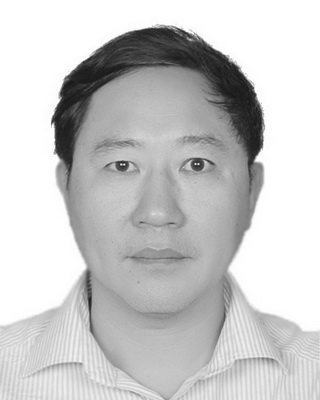}}]{Jiwu Huang (M'98--SM'00--F'16) }
  received the B.S. degree from Xidian University, Xian, China, in
  1982, the M.S. degree from Tsinghua University, Beijing, China, in
  1987, and the Ph.D. degree from the Institute of Automation, Chinese
  Academy of Sciences, Beijing, in 1998. He is currently a Professor
  with the College of Information Engineering, Shenzhen University,
  Shenzhen, China. His current research interests include multimedia
  forensics and security. He is a member IEEE Signal Processing
  Society Information Forensics and Security Technical Committee and
  serves as an Associate Editor for the IEEE Transactions on
  Information Forensics and Security. He was a General Co-Chair of the
  IEEE Workshop on Information Forensics and Security in 2013 and a
  TPC Co-Chair of the IEEE Workshop on Information Forensics and
  Security in 2018.
\end{IEEEbiography}

\vfill
\end{document}